\RequirePackage{fix-cm}

\documentclass[smallcondensed,envcountsect]{article}       
\usepackage[dvipsnames]{xcolor}
\usepackage{graphicx}
\usepackage{natbib}
\usepackage[multiple]{footmisc}
\setcitestyle{aysep={}} 
\usepackage[hidelinks]{hyperref}%
\usepackage{multirow}
\usepackage{tikz}
\usepackage{pgfplots}
\usepackage{subcaption} 
\usepackage{adjustbox}
\usepackage{bm}
\usepackage{mathtools}

\newtheorem{mydef2}{Definition}[section]{\bfseries}{\itshape}

\usepackage{array}
\makeatletter
\newcommand{\thickhline}{%
	\noalign {\ifnum 0=`}\fi \hrule height 1pt
	\futurelet \reserved@a \@xhline
}
\newcolumntype{"}{@{\hskip\tabcolsep\vrule width 1pt\hskip\tabcolsep}}
\makeatother

\newcolumntype{R}[2]{%
	>{\adjustbox{angle=#1,lap=\width-(#2)}\bgroup}%
	l%
	<{\egroup}%
}

\usepackage{amssymb}
\usepackage{pifont}

\begin{document}

\title{Temporal Network Comparison using\\
	Graphlet-orbit Transitions}



\author{David Apar\'icio \and Pedro Ribeiro \and Fernando Silva}




\maketitle

\begin{abstract}
	Networks are widely used to model real-world systems and uncover their topological features. Network properties such as the degree distribution and shortest path length have been computed in numerous real-world networks, and most of them have been shown to be both scale-free and small-world networks. Graphlets and network motifs are subgraph patterns that capture richer structural information than aforementioned global network properties, and these local features are often used for network comparison. However, past work on graphlets and network motifs is almost exclusively applicable only for static networks. Many systems are better represented as temporal networks which depict not only how a system was at a given stage but also how they evolved. Time-dependent information is crucial in temporal networks and, by disregarding that data, static methods can not achieve the best possible results. This paper introduces an extension of graphlets for temporal networks. Our proposed method enumerates all 4-node graphlet-orbits in each network-snapshot, building the corresponding orbit-transition matrix in the process. Our hypothesis is that networks representing similar systems have characteristic orbit transitions which better identify them than simple static patterns, and this is assessed on a set of real temporal networks split into categories. In order to perform temporal network comparison we put forward an orbit-transition-agreement metric (OTA). OTA correctly groups a set of temporal networks that both static network motifs and graphlets fail to do so adequately. Furthermore, our method produces interpretable results which we use to uncover characteristic orbit transitions, and that can be regarded as a network-fingerprint.

\end{abstract}

\section{Introduction}

Networks (or graphs) are widely used to model real-world systems, which are often too complex to be studied in their entirety. Networks represent entities of a system as nodes and their relations as edges connecting them (e.g. researchers are nodes in co-authorship networks and edges connect those that publish together). Networks are a useful representation for all kinds of social, biological and communication processes \citep{costa2011analyzing}. Modeling systems as networks is convenient not only for their conciseness, since they hide excessive detail of the original systems, but also because many network/graph properties are well-known. Degree distribution and shortest path length have shown that numerous real-world networks are both scale-free \citep{barabasi1999emergence} and small-world networks \citep{watts1998collective}. Brain networks in particular have been identified as small-world networks \citep{sporns2004organization}, meaning that each node (representing either a brain region in mesoscale conectomes or a neuron in microscale connectomes) is only a few connections away from any other node. Furthermore, the average path length in the brain has been negatively correlated with a person's IQ \citep{van2009efficiency}, further suggesting the correlation between small-world organization and network efficiency. Other properties such as the clustering coefficient and modularity are also frequently used to characterize a network.

Another approach to uncover the underlying structure of a network is to decompose it into smaller components (or subgraphs). Local network metrics such as network motifs \citep{milo2002network} and graphlets \citep{przulj2007} show that recurrent subgraph analysis often leads to a better insight into the network's function than global properties, due to the richer topological information that the former contain. Network motif analysis has identified the feed-forward loop as a recurring and crucial functional pattern in many real biological networks, such as gene regulation and metabolic networks \citep{mangan2003structure, zhu2005structural}. Graphlet-degree-agreement is a useful metric for network comparison and model fitting of real networks. Graphlets were initially proposed by \cite{przulj2007} to show that protein-protein interaction networks are more akin to geometric graphs than with traditional scale-free models. 

While temporal extensions for network motifs and graphlets have been put forward (which are discussed in Section~\ref{sec:relatedwork}), they were initially proposed for static networks, and have been mostly limited to them. Static networks disregard when edges occur, and instead aggregate all temporal information into a single final state.  This reduction is often deficient since it might be more telling to analyze how a specific network \emph{evolved}, rather than how it \emph{is} at an isolated stage. Previous studies have concluded that cliques and near-cliques are overrepresented in co-authorship networks \citep{choobdar2012comparison, pan2012strength}. However, these studies analyzed only final aggregate networks, which are much denser than network snapshots and not representative of the actual system. Moreover, such studies do not offer any intuition on how the network reached that state, how stable cliques are or which patterns are more unstable. 

This work puts forward a methodology for temporal network interpretation and comparison. Our method begins by enumerating transitions between all $k$-node graphlet-orbits in network $N$. Orbits are enumerated instead of simple graphlets since more information is obtained by counting the former (e.g. nodes at the periphery of a star are structurally distinct from the star-center, and exchanging places should be accounted for). Orbit-transition matrices $\mathcal{T}_{N_i}$ are built for each network $N_i$. These matrices are compared using our orbit-transition-agreement metric (OTA) in order to assess network similarity. Orbit-transitions were enumerated for more than a dozen real-world temporal networks pertaining to different categories, such as collaboration, crime, email communication, physical interaction and bipartite networks. Our method is capable of grouping similar themed networks whereas traditional static network motifs and graphlets can not.

The remainder of this paper is organized as follows. First, Section~\ref{sec:relatedwork} presents related work on network comparison using subgraph-based metrics, both for static and temporal networks. Next, Section~\ref{sec:compsim} begins by introducing necessary graph terminology and then presenting our proposed method for temporal network comparison. Experimental results are discussed in Section~\ref{sec:exp}, where various metrics are used to compare and group a set of networks from different categories and interpret the results. Finally, Section~\ref{sec:conclusions} presents the main conclusions of this work.

\section{Related work}\label{sec:relatedwork}

Performing network comparison is often useful, particularly because if properties of a given network are well known it allows for knowledge transfer to similar networks \citep{kelley2003conserved}. Global metrics such as degree distribution, characteristic path length and clustering coefficient give an idea of the structure of the networks and can be used to compare them. For instance, social networks tend to have an higher clustering coefficient and a smaller characteristic path length than spatial networks. Subgraph-based metrics offer richer topological information than simple global networks, and we discuss them in this section. Previous approaches have extended motifs \citep{jin2007trend, kovanen2011temporal} and graphlets \citep{hulovatyy2015exploring} to temporal networks; their differences in regards to our method are also presented in this section. 

\subsection{Static subgraphs}

Network motif fingerprints \citep{milo2002network} and graphlet-based metrics \citep{przulj2007} have been used for network comparison \citep{milo2004superfamilies, aparicio2016extending}, and both approaches compute the frequency of small non-isomorphic subgraphs (see Definiton~\ref{def:genproblem}). Graphlets evaluate the contribution of each individual node from the network, producing a graphlet degree distribution that can be seen as an extension of the node degree concept. Enumerating network motifs is computationally expensive since subgraphs are not only enumerated in the original network but also on a large set of similar randomized networks in order to assess motif significance. \cite{milo2004superfamilies} compared network motifs with three and four nodes of four superfamilies: sensory networks, hyperlink networks, social networks and linguistic networks. By comparing motif significances they were able to correctly cluster all four superfamilies. Similar studies have been carried out to classify metabolic networks \citep{zhu2005structural}, co-authorship networks \citep{choobdar2012comparison} or articles \citep{wu2012classifying}. Another possibility is to, instead of directly comparing two real networks, compare a network with models. \cite{przulj2007} showed that protein-protein interaction networks were more accurately described as random geometric graphs rather than purely random or scale-free networks. Therefore, motifs and graphlets have been successfully used to compare static networks. However, metrics such as these disregard temporal information which can be crucial for a better understanding of network topology and function.

\subsection{Dynamic subgraphs}

It is often more meaningful to evaluate how a network evolved through time rather than how the network is in its final state. Enumerating all possible subgraphs of a certain size $k$ is very computationally demanding, therefore some previous work has selected small subgraphs or only a few larger ones. Triangles are meaningful for many applications since they are one of the simplest forms of community. \cite{buriol2006counting} and  \cite{pavan2013counting} put forward a method to extract approximate and exact counts of all triangles in graph streaming environments. \cite{finocchi2014counting} proposed an algorithm to count cliques for $k$ slightly larger than 3. Instead of triangles, \cite{aliakbarpour2016sublinear} focused on $k-$star graphs.

\cite{kovanen2011temporal} presented an extension of network motifs to event networks and studied them on a phone call network. Their proposed temporal motifs have at most three events and a varying number of nodes. A similar approach for graphlets was put forward by \cite{hulovatyy2015exploring}; their dynamic graphlets have a fixed number of events (3 or 4) and a varying number of nodes. Both temporal motifs and dynamic graphlets were shown to be more reliable for network classification than static measures. However, their approach only allows for one event at a time. As a consequence, these methods do not capture situations where a loosely connected subgraph immediately becomes a clique or near-clique and are not applicable to event networks that intrinsically have multiple events occurring at the same time (i.e. when three authors collaborate on the same paper).

\cite{martin2016graphlet} proposed a metric to evaluate network similarity based on how their triplets are evolving over time. Their metric is based on the loss or gain of edges from one state to the next. Our method differs from theirs since they do not differentiate by pair-wise graphlet transitions but only by increase or decrease of total edges between states (i.e. different pair-wise transitions are not differentiated as long as they affect the same number of edges). The approach by \cite{doroud2011evolution} is more similar to our own since they enumerate all transitions between \mbox{3-node} directed subgraphs in network snapshots. That information is used in order to estimate the probability of a given transition in a social network and predict network changes. \cite{kim2012spatiotemporal} also count all \mbox{3-node} directed subgraphs to assess which motifs are present in different states of developing gene networks in different regions. These approaches are limited to a single network category (social network and gene expression networks, respectively), to 3-node subgraphs, they do not consider orbits and are not used for network comparison.


\section{Computing similarity using evolving graphlets}\label{sec:compsim}

This sections begins by introducing graph terminology used throughout this work and presenting the concept of static graphlets and graphlet enumeration. This is followed by a definition of temporal networks. Finally, evolving graphlets are put forward as well as our related metrics designed for network comparison.

\subsection{Graph terminology}

A network or \textit{graph} $G$ is comprised of a set $V(G)$ of \textit{vertices} or \textit{nodes} and a set $E(G)$ of \textit{edges} or \textit{connections}. Nodes represent entities and edges correspond to relationships between them. Edges are represented as pairs of vertices of the form $(a, b)$, where $a, b \in V(G)$. In \textit{directed} graphs, edges $(i, j)$ are \textit{ordered pairs} (translated to "$i$ \textit{goes to} $j$") whereas in \textit{undirected} graphs there is no order since the nodes are always reciprocally connected. 

A $subgraph$ $G_k$ of $G$ is a graph of size $k$ where $V(G_k) \subseteq V(G)$ and $E(G_k) \subseteq E(G)$. A subgraph is $induced$ if $\forall u, v \in V(G_k): (u,v) \in E(G_k)$ iff $(u, v) \in E(G)$. A $match$ or $occurrence$ of $G_k$ happens when $G$ has a set of nodes that induce $G_k$. Two matches are considered distinct if they have at least one different vertex. The $frequency$ of $G_k$ in $G$ is the number of occurrences of $G_k$ in $G$.

Two graphs are said to be \emph{isomorphic} if it is possible to obtain one from the other just by changing the node labels without affecting their topology. All occurrences of a set $\mathcal{G}$ of non-isomorphic subgraphs must be enumerated in the original network before graphlet or network motif metrics can be calculated. We call this task the general subgraph census problem \citep{wasserman1994social} and state it in Definition~\ref{def:genproblem}.

\begin{mydef2}[\textbf{Subgraph Census}]
	\label{def:genproblem}
	Given a set $\mathcal{G}$ of non-isomorphic subgraphs and a graph $G$, determine the frequency of all induced occurrences of the subgraphs $G_s \in \mathcal{G}$ in $G$. Two occurrences are considered different if they have at least one node or edge that they do not share. Other nodes and edges can overlap.
\end{mydef2}

\subsection{Static graphlets}\label{sec:staticgraphlets}

\begin{figure}
	\centering
	\includegraphics[width=0.7\linewidth]{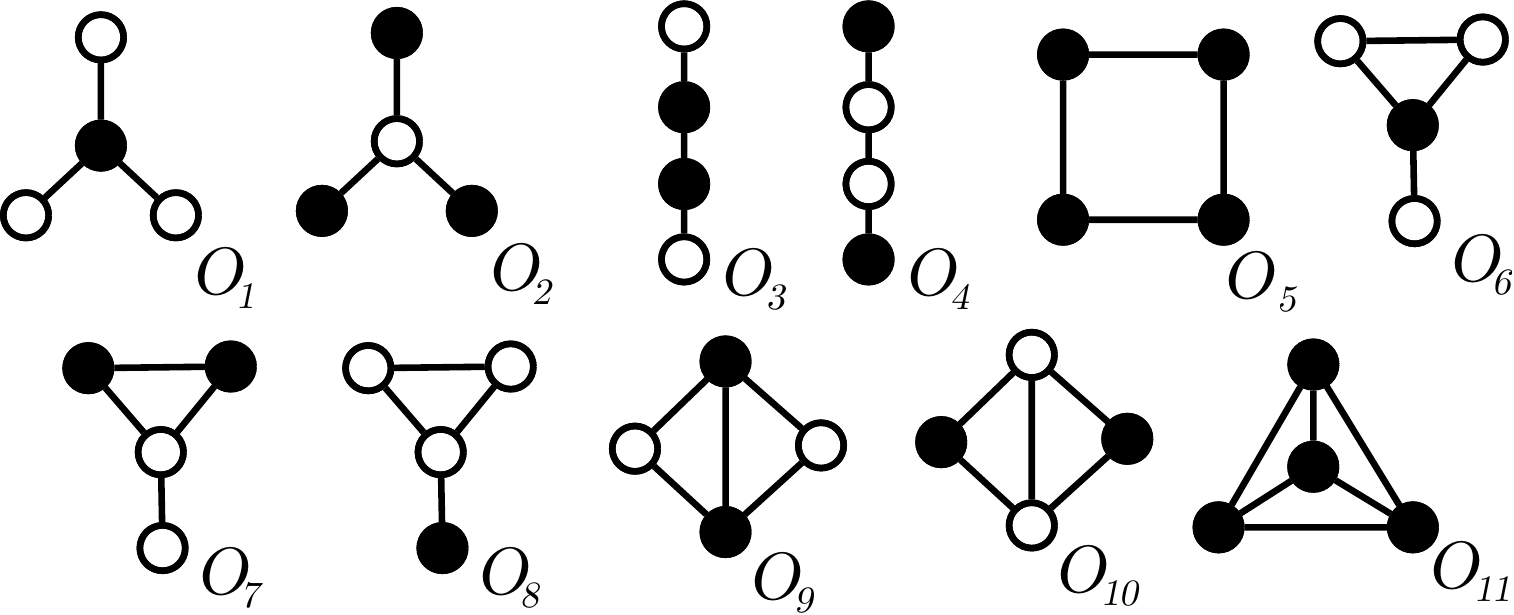}
	\caption{Set of all 11 graphlet-orbits of subgraphs with 4 nodes: $u\mathcal{O}_4$. Black nodes represent nodes at the orbit being considered.}
	\label{fig:orbits_u4}
\end{figure}

Graphlets \citep{przulj2007} are small induced non-isomorphic subgraphs that differentiate nodes according to their subgraph position, also called \emph{orbit}. For instance, the single node at the center of a \emph{star} is topologically different from a leaf-node, whereas leaf-nodes are structurally equivalent. Therefore, a \mbox{\emph{k-star}} has only two orbits: a \emph{center-orbit} which a single node inhabits and a \emph{leaf-orbit} where the remaining $k-1$ nodes are at. Graphlets can be either undirected~\citep{przulj2007} or directed subgraphs~\citep{aparicio2016extending}. Notation $u\mathcal{G}_k$ is adopted for the set of all undirected graphlets with $k$ nodes, and $d\mathcal{G}_k$ for the directed equivalent. The set of all orbits of $u\mathcal{G}_k$ is expressed as $u\mathcal{O}_k$, and $d\mathcal{O}_k$ is used for directed graphlets. For simplicity, prefixes $d$ and $u$ are suppressed whenever concepts are applicable to both directed and undirected graphlets. Figure~\ref{fig:orbits_u4} presents all undirected graphlet-orbits with 4 nodes.

\begin{figure}[h]
	\centering
	\includegraphics[width=1.0\linewidth]{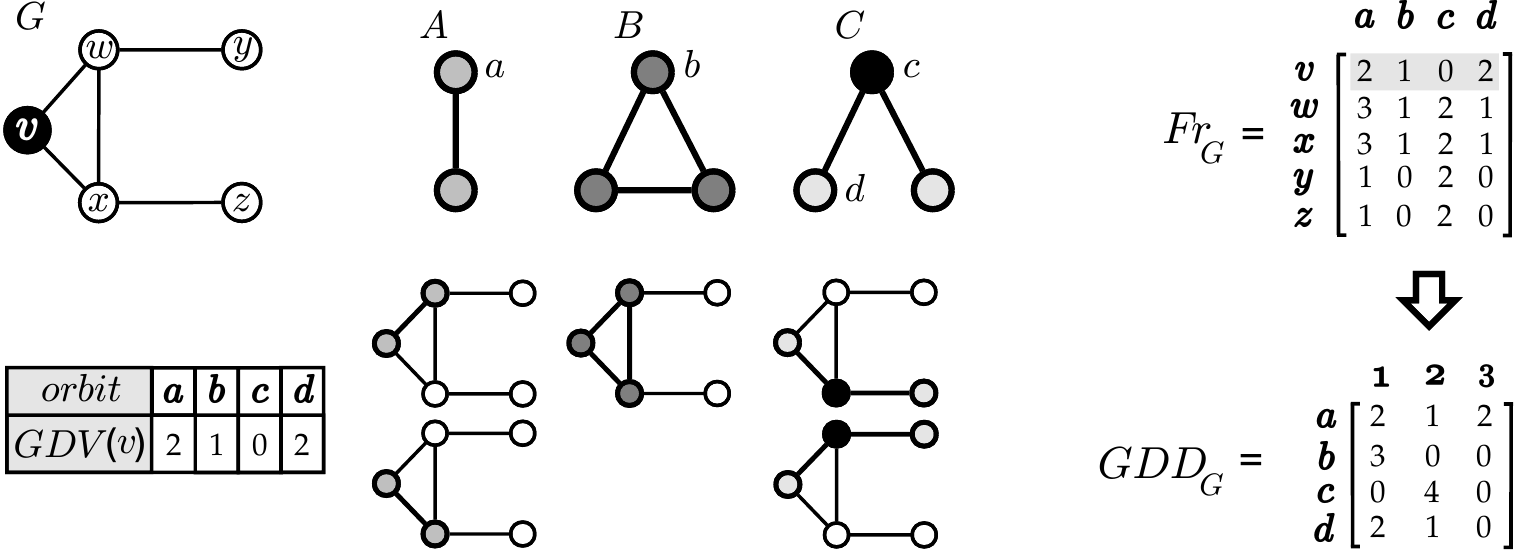}
	\caption{$GDV(v)$ obtained by enumerating the induced occurrences of all undirected graphlet orbits of sizes 2 and 3 ($A$, $B$ and $C$) touching $v$, and resulting $Fr(G)$ and $GDD(G)$ matrices for the complete subgraph census ($GDV(v)$ is highlighted in gray in $Fr(G)$.}
	\label{fig:orbit_occ}
\end{figure}

Graphlet-degree distributions are an extension of the node degree-distribution and both can be used for network comparison. In order to compute the degree distribution of a given graph $G$ one has to count $\forall u \in V(G)$ how many direct connections it has, also called node-degree. This task produces a vector of size $n$ containing the degrees of each $u \in V(G)$ which is transformed into a vector $Fr$ of size $m$, where $m$ is the maximum degree, and $Fr_p$ is the number of nodes that have degree $p$. Graphlet degree distributions generalize this concept for subgraphs bigger than the degree (actually, the degree is $u\mathcal{G}_2$). To compute the graphlet degree distribution it is necessary to count $\forall u \in V (G)$ how many times $u$ appears in some orbit $j \in \mathcal{O}$ and repeat this process for the total $o$ orbits, resulting in a graphlet degree vector $GDV(u)$ with $m$ positions.  A matrix $F r(G)$ of $n \times m$ positions is obtained by joining the $GDV$s of all $n$ nodes where each row of $Fr(G)$ is $GDV(v), v \in V(G)$ and each position $fr_{u,j}$ is the number of times that node $u$ appears in orbit $j$. This task is more formally defined in Definition~\ref{def:genproblem2} and an example of this process is given in Figure~\ref{fig:orbit_occ}. 

Two networks $G$ and $H$ are compared by computing the differences between their respective $GDD$ matrices after their distributions are normalized -- represented below as $n^j_G(k)$. One possibility to compare the two matrices is to use the arithmetic mean GDD-agreement ($GDA$) introduced by \cite{przulj2007}. A high $GDA(G,H)$ means that $G$ and $H$ are topologically similar; the $GDA(G, H)$ is defined as follows.

\begin{equation}
GDA(G,H)^j = 1 - \frac{1}{\sqrt{2}} \sqrt{\left(\sum\limits_{k=1}^{{\!+}\infty}[n^j_G(k) - n^j_H(k)]^2\right)}
\end{equation}

\begin{equation}
GDA(G,H) = \frac{1}{m}\sum\limits_{j=0}^{m}GDA(G,H)^j
\end{equation}

\begin{mydef2}[\textbf{Graphlet-Orbit Frequency Computation}]
	\label{def:genproblem2}
	Given a set $\mathcal{G}_s$ of non-isomorphic subgraphs of size $s$ and a graph $G$, determine the number of times $fr_{i,j}$ that each node $i \in V(G)$ appears in all the orbits $j \in \mathcal{O}_s$. All occurrences are induced. Two occurrences are considered different if they have at least one node or edge that they do not share. Other nodes and edges can overlap.
\end{mydef2}

Graphlets are widely used to analyze and compare static networks due to their rich topological information. However, analyzing static networks only gives information about a single state of the network, thus disregarding their evolution. Studying network changes is crucial in several networks: (i) in brain networks it is important to know which brain regions where activated in sequence to establish functional connections, (ii) some chains of events in event networks (such as bank transactions) might be benign or suspicious, \mbox{(iii) in} collaboration networks it is important not only to know that two authors collaborated but also when and if that collaboration was maintained through time. \cite{choobdar2012comparison} concluded that, according to their motif representation, chemists and physicists establish stronger groups than computer scientists or engineers. While this is an interesting conclusion, more relevant questions can be asked when considering temporal information (which are not limited to collaborations): is the group structure stable or does it change? How does it change? Are tightly connected groups more or less stable? How long does it take to go from a loosely connected group to a tightly connected one? What are the differences between temporal networks of different kinds? Temporal network definitions are given in the following section, as well as our method to answer the aforementioned questions. 



\subsection{Temporal network and evolving graphlets terminology}

\subsubsection{Temporal networks}
\label{sec:tempnets}

Temporal networks used throughout this paper consist of $s$ consecutive snapshots of a global network $G$. The set of all snapshots of $G$ is referred to as $\mathcal{S}$. An edge $(u, v)$ exists in $\mathcal{S}_i$ if nodes $u$ and $v$ are connected in the interval $[I + t \times i, I + t \times (i+1)[$, where $I$ is the starting time of the network (i.e. January 2000). Parameters $t$ and $s$ may be different depending on the network; for instance, in scientific co-authorship networks one or two years are the more suitable value for $t$, while in conference interaction networks $t$ is a few hours or a couple of days. The number of snapshots $s$ depends on the amount of available data.

Networks considered in this work can gain or lose edges and nodes from $\mathcal{S}_i$ to $\mathcal{S}_{i+1}$. Sometimes it might be useful for edges to be \emph{permanent} meaning that when they are added in $\mathcal{S}_i$ they remain in the network for all $\mathcal{S}_{i'}, i' >  i$. These are referred to as \emph{aggregate networks}. Whenever and edge from $\mathcal{S}_i$ must be activated to also be present in $\mathcal{S}_{i+1}$ the network is said to be an \emph{active-edge network}. Using aggregate or active-edge representations depends on the study being performed; for instance, if one wants to analyze how a scientific community is growing it might be more suitable to see how the aggregate network is evolving (are new people joining the community?) since authors do not have to publish a paper together every year to be considered a community. On the other hand, active edges should be considered if one wants to assess how stable a certain community is.	

\subsubsection{Evolving graphlets}

Only connected graphlets are taken into account in this work because our focus is to study how groups evolve and, when a group becomes disconnected, it is arguable if it is still a group. We should point out that disconnected graphlets would be very useful to analyze group formation. However, subgraph enumeration itself is a known NP-Complete problem and, in the worst case, enumerating all possible connected and disconnected graphlets would have complexity $\mathcal{O}(n^k)$, which is only feasible for small networks and very small $k-$graphlets.

\begin{figure}[h]
	\centering
	\includegraphics[width=1.0\linewidth]{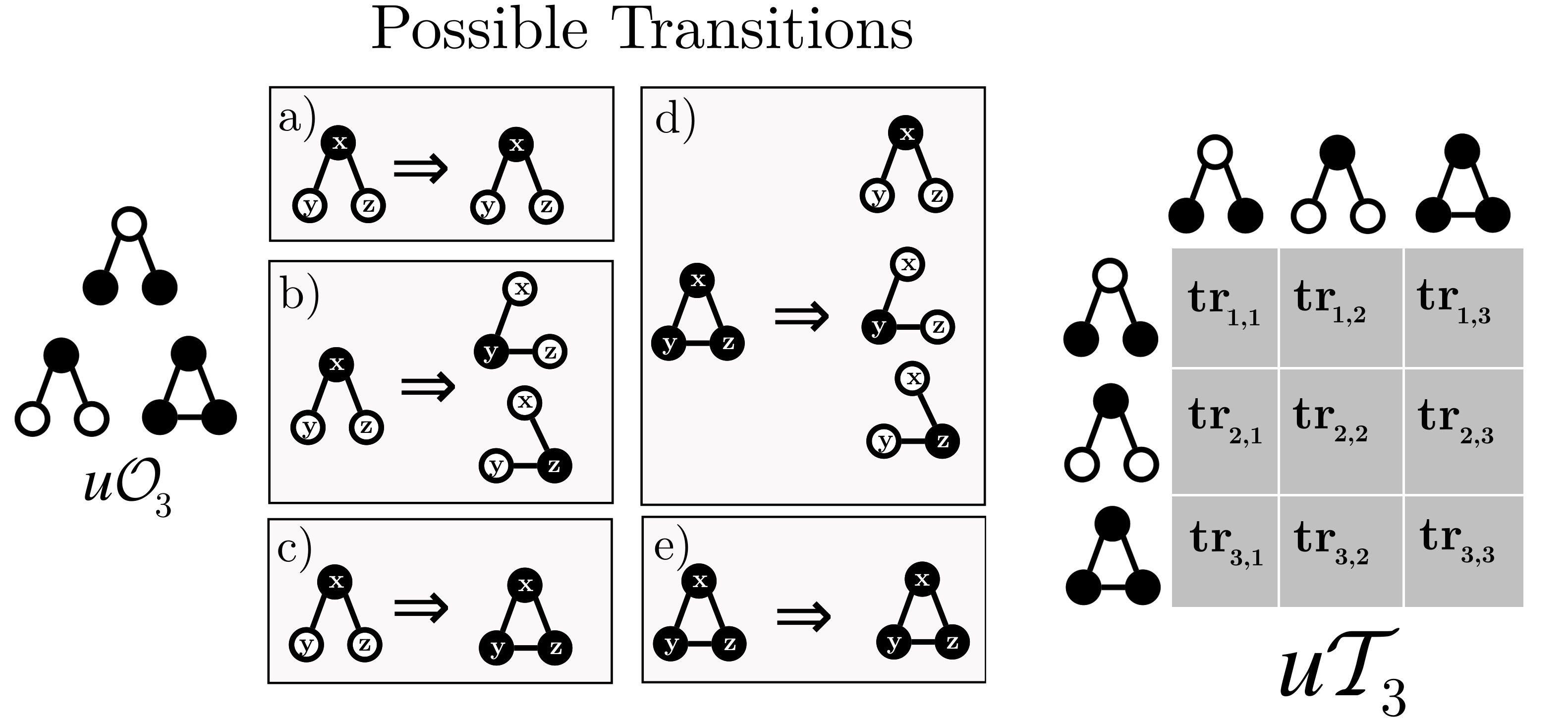}
	\caption{All possible orbit transitions of 3-node undirected graphlets ($u\mathcal{O}_3$) and corresponding orbit-transition matrix ($u\mathcal{T}_3$).}
	\label{fig:transitions_u3}
\end{figure}

In order to compare temporal networks' topology we evaluate how similar their graphlet-orbit transitions are. Consider the two possible 3-node undirected graphlets, $u\mathcal{G}_3$, and their respective orbits, $u\mathcal{O}_3$, from Figure~\ref{fig:transitions_u3}. The chain-graph has two possible positions -- the node can be either at its center (orbit-2) or in one of its leaves (orbit-1) -- while all nodes in a triangle-graph are topologically equivalent (orbit-3). At the subgraph-level there are 4 possible transitions: a) a chain remains as a chain, c) a chain becomes a triangle, d) a triangle breaks into a chain and e) a triangle stays as a triangle. Considering orbit-level transitions adds \mbox{b) chain-rotations} and differentiates nodes that go from a triangle-position (orbit-3) to a chain-center (orbit-2) or to a chain-leaf (orbit-1). Our algorithm enumerates these transitions and stores them in matrix $u\mathcal{T}_k$. For instance, when \mbox{d) occurs} one node goes from orbit-3 to orbit-2 ($tr_{3,2}$) while the other two transition from orbit-3 to orbit-1 ($tr_{3,1}$). Orbit-level transitions offer more information than subgraph-level transitions, therefore the former are used in this work. Matrix $u\mathcal{T}_3$ stores the orbit-transition frequencies of $u\mathcal{O}_3$. These matrices offer rich topological information that can be used for network summarization, Data Mining (they can be used as features for prediction tasks), network comparison and model fitting. In this work we compare different networks according to their transition matrices. Next we describe our metric for network comparison based on orbit-transition matrices. 

\subsubsection{Orbit temporal agreement (OTA)}
\label{sec:ota}

After enumerating all graphlet-orbit transitions, and having constructed $\mathcal{T}_k$ matrices for each network of set $\mathcal{N}$, our method computes their topological similarity. Orbits of size $k$ are enumerated for each network, therefore all $\mathcal{T}_k$ matrices consist of $|\mathcal{O}| \times |\mathcal{O}|$ transitions. Our approach is based on the arithmetic mean of orbit-transition differences. Matrices $\mathcal{T}_k$ are normalized before computing orbit-transitions differences in order to reduce bias induced by different network sizes. Normalization is performed by row, as shown in Equation~\ref{eq:norm}. This choice gives the same importance to common and rare orbits. Instead, one could normalize the matrix both by row and column if the scale of the original values is important. We feel that choosing the latter option would disregard differences in rare orbits, which can differentiate networks better than common ones.

\begin{equation}\label{eq:norm}
ntr_{i,j} = \frac{tr_{i,j}}{\sum\limits_{k=1}^{|\mathcal{O}|}tr_{i,k}}
\end{equation}

The similarity of two networks $G_1$ and $G_2$ is given by the average similarity of their graphlet-transition frequency for each graphlet-transition $ntr_{i,j}$. Equation~\ref{eq:ota} presents this metric, which we name orbit-transition agreement ($OTA$).

\begin{equation}\label{eq:ota}
OTA(G_1, G_2) = \frac{1}{|\mathcal{O}|} \times \sum\limits_{i=1}^{|\mathcal{O}|}\sum\limits_{j=1}^{|\mathcal{O}|}\Big(1-|ntr_{i,j}^{G_1} - ntr_{i,j}^{G_2}|\Big)
\end{equation}

Equation~\ref{eq:ota} produces an \emph{absolute value of agreement}: $OTA(G_1, G_2)$ is always the same regardless of $\mathcal{N}$. However, for our purposes a \emph{relative value of agreement} is more suitable since we want to compare networks within a specific set. For instance, some transitions are not possible in aggregate networks (such as a triangle becoming a chain) or always true (such as a clique remaining a clique). These transitions should not be taken into account for network similarity since they are intrinsic to the particular type of network. Furthermore, some transitions might be very frequent while others are rare. Each $\mathcal{T}_k$ matrix is normalized in respect to the minimum and maximum values found for set $\mathcal{N}$ before $OTA(G_i, G_j)$ is computed, as shown in Equation~\ref{eq:norm2}.

\begin{equation}\label{eq:norm2}
rntr_{i, j} = \frac{ntr_{i, j} - min(ntr_{i, j}, \mathcal{N})}{max(ntr_{i,j}, \mathcal{N}) - min(ntr_{i,j}, \mathcal{N})}
\end{equation}

\section{Experimental results}\label{sec:exp}

In this section we show the effectiveness of our proposed method in grouping a set of temporal networks by predetermined categories and identifying their similarities. The set of real-world networks $\mathcal{N}$ comprises (i)~co-authorship, (ii)~crime, (iii)~e-mail communication, (iv)~physical interaction, (v)~bipartite, (vi)~soccer transfers and (vii)~social media friendship networks, as shown on Table~\ref{tab:nets}. Our hypothesis is that networks of the same category have similar topological structure, and this is verified by our method. We begin by analyzing how networks are evolving over time (\emph{growing} vs. \emph{shrinking}, becoming \emph{more-connected} vs. \emph{less-connected}) as well as some of their global metrics, namely the \emph{average-degree}, the \emph{clustering-coefficient} and the \emph{characteristic path-length}. Network motif and graphlet analyses are also conducted since they capture richer topological information than aforementioned global metrics. We compare the networks' \emph{motif-fingerprints} and \emph{graphlet-degree distributions} for 4-node subgraphs and assess how well the networks are being grouped using these metrics. In order to compare the clustering capabilities of static graphlet-orbits with evolving graphlet-orbits we compute the graphlet-degree-agreement ($GDA$) and orbit-transition-agreement ($OTA$) for each pair of networks and cluster the set $\mathcal{N}$ accordingly: networks with high agreement are grouped together. Finally, we show that graphlet-orbit transition matrices offer highly interpretable information which displays both (a) clear differences between networks of different categories and (b) characteristic transitions in networks of the same category.

\subsection{Network overview}

\begin{table}
	\scriptsize
	\centering
	\def\arraystretch{1.0}
	\caption{Set of temporal networks $\mathcal{N}$ grouped by category. $\rho$ is the time-interval for each snapshot and ${|\mathcal{S}|}$ is the number of snapshots.}
	\label{tab:nets}
	\begin{tabular}{l p{1.4cm}@{ }@{}c@{}@{ }p{1.4cm} c l l l  }
		\thickhline 
		& & & & & & & \\[-5pt]
		{\bf Name} & \multicolumn{3}{c}{\bf Nodes} & {\bf Edges} & & $\boldsymbol{\rho}$ &  $\boldsymbol{|\mathcal{S}|}$ \\ [2pt]
		\hline
		& & & & & & & \\[-6pt]

		\multirow{2}{*}{\texttt{Authenticus}} & \multicolumn{3}{c}{authors}  & co-author a paper & & \multirow{2}{*}{1 year} & \multirow{2}{*}{16}   \\ 
		& \multicolumn{3}{c}{7k}  & 120k &  &  & \\[3pt] 
		
		\multirow{2}{*}{\texttt{arXiv hep-ph}}   & \multicolumn{3}{c}{authors}  & co-author a paper & & \multirow{2}{*}{1 year}  & \multirow{2}{*}{7}  \\
		& \multicolumn{3}{c}{2k}  & 357k &   & & \\ [2pt] \hline\\[-6pt]
		
		\multirow{2}{*}{\texttt{Minneapolis}}   & \multicolumn{3}{c}{streets}  & crime in intersection & & \multirow{2}{*}{3 months}  & \multirow{2}{*}{16}  \\
		& \multicolumn{3}{c}{454}    & 12k &   & & \\ [3pt]
		
		\multirow{2}{*}{\texttt{Philadelphia}}   & \multicolumn{3}{c}{streets}  & crime in intersection & & \multirow{2}{*}{3 months}  & \multirow{2}{*}{16}  \\
		& \multicolumn{3}{c}{1k}    & 10k &  & & \\ [2pt] \hline\\[-6pt] 
		
		\multirow{2}{*}{\texttt{Emails}}   & \multicolumn{3}{c}{workers}  & email between workers & & \multirow{2}{*}{1 month}  & \multirow{2}{*}{9}  \\
		& \multicolumn{3}{c}{167}    & 83k &  & & \\ [3pt]
		
		\multirow{2}{*}{\texttt{Enron}}   & \multicolumn{3}{c}{workers}  & email between workers & & \multirow{2}{*}{2 months}  & \multirow{2}{*}{16}  \\
		& \multicolumn{3}{c}{6k}    & 51k &  & & \\ [2pt] \hline\\[-6pt]
		
		\multirow{2}{*}{\texttt{Gallery}}   & \multicolumn{3}{c}{visitors}  & physical interaction & & \multirow{2}{*}{4 days}  & \multirow{2}{*}{16}  \\
		& \multicolumn{3}{c}{420}    & 43k &  & & \\ [3pt]
		
		\multirow{2}{*}{\texttt{Conference}}   & \multicolumn{3}{c}{visitors}  & physical interaction & & \multirow{2}{*}{12 hours}  & \multirow{2}{*}{6}  \\
		& \multicolumn{3}{c}{113}    & 21k &  & & \\ [3pt]
		
		\multirow{2}{*}{\texttt{School}}   & \multicolumn{3}{c}{students}  & physical interaction & & \multirow{2}{*}{1 day}  & \multirow{2}{*}{5}  \\
		& \multicolumn{3}{c}{327}    & 189k &  & & \\ [3pt]
		
		\multirow{2}{*}{\texttt{Workplace}}   & \multicolumn{3}{c}{workers}  & physical interaction & & \multirow{2}{*}{10 days}  & \multirow{2}{*}{10}  \\
		& \multicolumn{3}{c}{92}    & 10k &  & & \\ [2pt] \hline\\[-6pt]
		
		\multirow{2}{*}{\texttt{Escorts}}   & \multicolumn{1}{r@{ }}{clients} & + & escorts  & hires & & \multirow{2}{*}{3 months}  & \multirow{2}{*}{16}  \\
		& \multicolumn{1}{r@{ }}{10k} & + & 7k    & 51k &   & & \\ [3pt]
		
		\multirow{2}{*}{\texttt{Twitter}}   & \multicolumn{1}{r@{ }}{users} & + & hashtags  & user tweets hashtag  & & \multirow{2}{*}{3 months}  & \multirow{2}{*}{16}  \\
		& \multicolumn{1}{r@{ }}{12k} & + & 16k    & 327k &   & & \\ [2pt] \hline\\[-6pt]
		
		\multirow{2}{*}{\texttt{Transfers}}   & \multicolumn{3}{c}{soccer teams}  & player transfer & & \multirow{2}{*}{1 year}  & \multirow{2}{*}{16}  \\
		& \multicolumn{3}{c}{2k}    & 20k &  & & \\ [2pt] \hline\\[-6pt]
		
		\multirow{2}{*}{\texttt{Facebook}}   & \multicolumn{3}{c}{friends}  & posts on the other's wall & & \multirow{2}{*}{3 months}  & \multirow{2}{*}{16}  \\
		& \multicolumn{3}{c}{47k}    & 877k &  & & \\[2pt]
		
		
		\thickhline
	\end{tabular}
\end{table}

A set of 14 temporal networks $\mathcal{N}$ was collected from various sources\footnote{\url{http://konect.uni-koblenz.de/networks}}\footnote{\url{https://www.kaggle.com/mrisdal/minneapolis-incidents-crime}}\footnote{\url{https://www.kaggle.com/mchirico/philadelphiacrimedata}}\footnote{\url{https://www.cs.cmu.edu/~./enron/}}\footnote{\url{https://github.com/axknightroad}}\footnote{\url{http://www.sociopatterns.org}}\footnote{\url{http://apps.who.int/gho/data}}\footnote{\url{http://vlado.fmf.uni-lj.si/pub/networks/data}}\footnote{\url{http://www.dcc.fc.up.pt/~daparicio/networks}} in order to evaluate our method's efficiency. $\mathcal{N}$ is comprised of active-edge networks, meaning that edges are only present in the snapshot $S_i$ in which they appear at and need to be re-activated in subsequent snapshots (see section~\ref{sec:tempnets}). The number of snapshots $|\mathcal{S}_N|$ depends on the amount of available data of $N$. Long-term networks, such as co-authorship networks, have a bigger time-interval $\rho$ when compared with short-term event networks, such as physical interactions.

Figure~\ref{fig:growth} shows how the networks are evolving size-wise. Most of them are growing as time elapses. The fastest growing networks are \texttt{arXiv hep-ph}, \texttt{Twitter}, \texttt{Facebook} and \texttt{Enron}, which start at only $\approx10\%$ of their largest state, but \texttt{Enron} begins shrinking at $t=11$ until it almost disappears. \texttt{Authenticus}, \texttt{Escorts} and \texttt{Transfers} are also growing networks but they grow at a slower rate and become almost stagnant at the end, where they might have reached their full potential in terms of growth. Crime, physical interaction networks and \texttt{Emails} stay relatively stable in size. Figure~\ref{fig:avgdegree_evol} presents the evolution of the networks' average degree. \texttt{arXiv hep-ph}, \texttt{Emails} and physical interaction networks are the ones with higher average degree. \texttt{arXiv hep-ph}, \texttt{Twitter} and \texttt{Facebook} are the fastest growing in terms of their average degree and most networks have a stable average degree. By observing Figure~\ref{fig:charpathlen_evol} one can conclude that all networks from $\mathcal{N}$ are small-world since their characteristic-path-length at latter stages ($t \approx 16$) is between 2 and 7. No correlation linking category with characteristic-path-length evolution exist, nor with growth or average degree. Clustering coefficients were also computed for each network snapshot and it was found that they do not change with $t$. Co-authorship networks have the highest clustering coefficient at 0.5 while crime, bipartite, \texttt{Facebook} and \texttt{Tranfers} networks have near-zero clustering coefficient. The clustering coefficient is capable of grouping co-authorship networks together despite only considering 3-node subgraphs (triangles and 3-node chains). However, it does not distinguish between crime and bipartite networks, for instance. In these cases, one option to differentiate between networks with similar 3-node subgraphs is to analyze their 4-node network motifs and graphlets, and this approach is followed in Sections~\ref{sec:motifs} and \ref{sec:graphlets}, respectively.

\pgfplotsset{every tick label/.append style={font=\scriptsize}}
\begin{figure}
	
	\centering
	\hspace{-1.2cm}
	\begin{subfigure}{.30\linewidth}\centering
		\begin{tikzpicture}
		\begin{axis}[
		scale only axis,
		y label style={at={(axis description cs:0.15,.5)}},
		width=0.6\linewidth,
		xlabel={{\bf \emph{t}:}},
		xmin=0, xmax=17,
		xtick={1,6,11,16},
		xmajorgrids,
		ymin = -.1, ymax=1.1,
		ylabel={\bf Relative Size},
		every axis x label/.style={
			at={(-0.05,-0.095)}
		},
		ymajorgrids,
		axis lines*=left,
		legend style ={ at={(0.47,1.26)}, 
			legend columns=2,
			font=\scriptsize,
			only marks,
			/tikz/every even column/.append style={column sep=0.2cm},
			anchor=north , draw=none, 
			fill=white,align=left},
		cycle list name=black white,
		smooth
		]
		
		\addplot[mark=square*,mark options={solid, fill=red},mark size=1.2] coordinates{
			(1,0.2481752)
			(2,0.2695696)
			(3,0.3143720)
			(4,0.3606846)
			(5,0.4465140)
			(6,0.4950919)
			(7,0.5753838)
			(8,0.6342814)
			(9,0.7274100)
			(10,0.8016612)
			(11,0.8489806)
			(12,0.9063680)
			(13,0.9831362)
			(14,0.9967279)
			(15,1.0000000)
			(16,0.9368236)
		};
		\addlegendentry{Authenticus};
		
		\addplot[,mark=*,mark options={solid, fill=red},mark size=1.2] coordinates{
			(10,0.06950673)
			(11,0.22242152)
			(12,0.39955157)
			(13,0.57892377)
			(14,0.71031390)
			(15,0.84573991)
			(16,1.00000000)
		};
		\addlegendentry{arxiv};
		\end{axis}
		\end{tikzpicture}%
	\end{subfigure}
	\begin{subfigure}{.30\linewidth}\centering
		\begin{tikzpicture}
		\begin{axis}[
		y label style={at={(axis description cs:0.05,.5)}},
		x label style={at={(axis description cs:0.5,.08)}},
		legend image post style={scale=0.8},
		scale only axis,
		width=0.6\linewidth,
		xtick={1,6,11,16},
		xmin=0, xmax=17,
		xmajorgrids,
		ymin = -.1, ymax=1.1,
		ylabel={\textcolor{white}{.}},
		ymajorgrids,
		axis lines*=left,
		xlabel={{\bf \emph{t}:}},
		every axis x label/.style={
			at={(-0.05,-0.095)}
		},
		legend style ={ at={(0.47,1.26)}, 
			legend columns=2,
			font=\scriptsize,
			only marks,
			/tikz/every even column/.append style={column sep=0.2cm},
			anchor=north , draw=none, 
			fill=white,align=left},
		cycle list name=black white,
		smooth
		]
		
		\addplot[mark=square*,mark options={solid, fill=blue}] coordinates{
			(1,0.8015267)
			(2,0.9236641)
			(3,0.9961832)
			(4,0.9236641)
			(5,0.7633588)
			(6,0.8778626)
			(7,0.9732824)
			(8,0.9427481)
			(9,0.7748092)
			(10,0.9580153)
			(11,1.0000000)
			(12,0.8358779)
			(13,0.7519084)
			(14,0.8625954)
			(15,0.9694656)
			(16,0.8244275)
		};
		\addlegendentry{Minneapolis};
		
		\addplot[mark=*,mark options={solid, fill=blue}] coordinates{
			(1,0.9323770)
			(2,1.0000000)
			(3,0.9200820)
			(4,0.8422131)
			(5,0.7581967)
			(6,0.8463115)
			(7,0.8176230)
			(8,0.7438525)
			(9,0.6168033)
			(10,0.7684426)
			(11,0.7274590)
			(12,0.6721311)
			(13,0.5655738)
			(14,0.7336066)
			(15,0.7254098)
			(16,0.6270492)
		};
		\addlegendentry{Philadelphia};
		\end{axis}
		\end{tikzpicture}%
	\end{subfigure}
	\begin{subfigure}{.30\linewidth}\centering
		\begin{tikzpicture}
		\begin{axis}[
		x label style={at={(axis description cs:0.5,.08)}},
		scale only axis,
		ymajorticks=false,
		width=0.6\linewidth,
		xtick={1,6,11,16},
		xmin=0, xmax=17,
		legend image post style={scale=0.8},
		xmajorgrids,
		ymin = -.1, ymax=1.1,
		ylabel={\textcolor{white}{.}},
		ymajorgrids,
		axis lines*=left,
		xlabel={{\bf \emph{t}:}},
		every axis x label/.style={
			at={(-0.05,-0.095)}
		},
		ymajorticks=true,
		legend style ={ at={(0.47,1.26)}, 
			legend columns=2,
			font=\scriptsize,
			only marks,
			/tikz/every even column/.append style={column sep=0.2cm},
			anchor=north , draw=none, 
			fill=white,align=left},
		cycle list name=black white,
		smooth
		]
		
		\addplot[mark=square*,mark options={solid, fill=violet}] coordinates{
			(8,1.0000000)
			(9,0.8741722)
			(10,0.9271523)
			(11,0.9668874)
			(12,0.9602649)
			(13,0.8807947)
			(14,0.8609272)
			(15,0.8874172)
			(16,0.9205298)
		};
		\addlegendentry{Emails};
		
		\addplot[mark=*,mark options={solid, fill=violet}] coordinates{
			(1,0.01691989)
			(2,0.08287293)
			(3,0.10704420)
			(4,0.11982044)
			(5,0.22306630)
			(6,0.44129834)
			(7,0.48964088)
			(8,0.62327348)
			(9,0.71132597)
			(10,0.79972376)
			(11,1.00000000)
			(12,0.80110497)
			(13,0.77313536)
			(14,0.64053867)
			(15,0.10531768)
			(16,0.01035912)
		};
		\addlegendentry{Enron};
		\end{axis}
		\end{tikzpicture}%
	\end{subfigure}
	
	\hspace{-1.2cm}
	\begin{subfigure}{.30\linewidth}\centering
		\vspace{0.08cm}
		\begin{tikzpicture}
		\begin{axis}[
		scale only axis,
		y label style={at={(axis description cs:0.15,.5)}},
		width=0.6\linewidth,
		xlabel={{\bf \emph{t}:}},
		xmin=0, xmax=17,
		xtick={1,6,11,16},
		xmajorgrids,
		ymin = -.1, ymax=1.1,
		ylabel={\bf Relative Size},
		every axis x label/.style={
			at={(-0.06,-0.07)}
		},
		legend image post style={scale=0.8},
		ymajorgrids,
		axis lines*=left,
		xlabel={{\bf \emph{t}:}},
		every axis x label/.style={
			at={(-0.05,-0.095)}
		},
		legend style ={ at={(0.47,1.46)}, 
			legend columns=2,
			font=\scriptsize,
			only marks,
			/tikz/every even column/.append style={column sep=0.2cm},
			anchor=north , draw=none, 
			fill=white,align=left},
		cycle list name=black white,
		smooth
		]
		
		\addplot[mark=square*,mark options={solid, fill=ForestGreen}] coordinates{
			(1,0.7357143)
			(2,0.5404762)
			(3,0.5904762)
			(4,0.4619048)
			(5,0.5761905)
			(6,0.3809524)
			(7,0.3142857)
			(8,0.3952381)
			(9,0.3333333)
			(10,0.3309524)
			(11,0.4095238)
			(12,0.2857143)
			(13,0.4357143)
			(14,0.5690476)
			(15,0.4785714)
			(16,1.0000000)
		};
		\addlegendentry{Gallery};
		
		\addplot[mark=*,mark options={solid, fill=ForestGreen}] coordinates{
			(11,0.8762887)
			(12,0.9896907)
			(13,1.0000000)
			(14,1.0000000)
			(15,0.9793814)
			(16,0.8969072)
		};
		\addlegendentry{Conference};
		
		\addplot[mark=triangle*,mark options={solid, fill=ForestGreen}] coordinates{
			(12,1.0000000)
			(13,0.9935897)
			(14,0.9711538)
			(15,0.9455128)
			(16,0.9583333)
		};
		\addlegendentry{School};
		
		\addplot[mark=diamond*,mark options={solid, fill=ForestGreen}] coordinates{
			(7,1.0000000)
			(8,0.9722222)
			(9,0.8194444)
			(10,0.9722222)
			(11,0.8611111)
			(12,0.9444444)
			(13,0.9583333)
			(14,0.9583333)
			(15,0.9444444)
			(16,0.8611111)
		};
		\addlegendentry{Workplace};
		\end{axis}
		\end{tikzpicture}%
	\end{subfigure}
	\begin{subfigure}{.30\linewidth}\centering
		\vspace{0.37cm}
		\begin{tikzpicture}
		\hspace{0.08cm}
		\begin{axis}[
		y label style={at={(axis description cs:0.05,.5)}},
		x label style={at={(axis description cs:0.5,.08)}},
		legend image post style={scale=0.8},
		scale only axis,
		width=0.6\linewidth,
		xtick={1,6,11,16},
		xmin=0, xmax=17,
		xmajorgrids,
		ymin = -.1, ymax=1.1,
		ylabel={\textcolor{white}{.}},
		ymajorgrids,
		axis lines*=left,
		ymajorticks=true,
		xlabel={{\bf \emph{t}:}},
		every axis x label/.style={
			at={(-0.05,-0.095)}
		},
		legend style ={ at={(0.47,1.26)}, 
			legend columns=2,
			font=\scriptsize,
			only marks,
			/tikz/every even column/.append style={column sep=0.2cm},
			anchor=north , draw=none, 
			fill=white,align=left},
		cycle list name=black white,
		smooth
		]
		
		\addplot[mark=square*,mark options={solid, fill=Dandelion}] coordinates{
			(5,0.009058883)
			(6,0.062405637)
			(7,0.196024157)
			(8,0.404378460)
			(9,0.538500252)
			(10,0.701560141)
			(11,0.821590337)
			(12,0.924509311)
			(13,0.952692501)
			(14,0.994967287)
			(15,1.000000000)
			(16,0.805988928)
		};
		\addlegendentry{Escorts};
		
		\addplot[mark=*,mark options={solid, fill=Dandelion}] coordinates{
			(8,0.006153846)
			(9,0.013002481)
			(10,0.041588089)
			(11,0.076575682)
			(12,0.130471464)
			(13,0.231166253)
			(14,0.387295285)
			(15,0.582928040)
			(16,1.000000000)
		};
		\addlegendentry{Twitter};
		\end{axis}
		\end{tikzpicture}%
	\end{subfigure}
	\begin{subfigure}{.30\linewidth}\centering
		\vspace{0.37cm}
		\begin{tikzpicture}
		\hspace{0.23cm}
		\begin{axis}[
		y label style={at={(axis description cs:0.05,.5)}},
		x label style={at={(axis description cs:0.5,.08)}},
		legend image post style={scale=0.8},
		ymajorticks=false,
		scale only axis,
		width=0.6\linewidth,
		xtick={1,6,11,16},
		xmin=0, xmax=17,
		xmajorgrids,
		ymin = -.1, ymax=1.1,
		ylabel={\textcolor{white}{.}},
		ymajorgrids,
		axis lines*=left,
		xlabel={{\bf \emph{t}:}},
		every axis x label/.style={
			at={(-0.05,-0.095)}
		},
		ymajorticks=true,
		legend style ={ at={(0.47,1.26)}, 
			legend columns=2,
			font=\scriptsize,
			only marks,
			/tikz/every even column/.append style={column sep=0.2cm},
			anchor=north , draw=none, 
			fill=white,align=left},
		cycle list name=black white,
		smooth
		]
		
		\addplot[,mark=square*,mark options={solid, fill=black}] coordinates{
			(1,0.5232404)
			(2,0.5325365)
			(3,0.5285525)
			(4,0.5604250)
			(5,0.6069057)
			(6,0.6759628)
			(7,0.7622842)
			(8,0.8844622)
			(9,0.9096946)
			(10,0.8579017)
			(11,0.9163347)
			(12,0.9402390)
			(13,0.9176627)
			(14,0.9521912)
			(15,0.9800797)
			(16,1.0000000)
			
		};
		\addlegendentry{Transfers};
		
		\addplot[,mark=*,mark options={solid, fill=SkyBlue}] coordinates{
			(1,0.004062637)
			(2,0.009230927)
			(3,0.013653545)
			(4,0.037797948)
			(5,0.058008280)
			(6,0.080918464)
			(7,0.111053971)
			(8,0.175619038)
			(9,0.221387982)
			(10,0.267105500)
			(11,0.323673858)
			(12,0.367771464)
			(13,0.375896737)
			(14,0.441361755)
			(15,0.564012239)
			(16,1.000000000)
			
		};
		\addlegendentry{Facebook};
		\end{axis}
		\end{tikzpicture}%
	\end{subfigure}
	\caption{Network growth according to its number of vertices -- grouped by type.} \label{fig:growth}
	
	\vspace{0.5cm}
	\hspace{-1.2cm}
	\begin{subfigure}{.30\linewidth}\centering
		\begin{tikzpicture}
		\begin{axis}[
		scale only axis,
		y label style={at={(axis description cs:0.15,.5)}},
		width=0.6\linewidth,
		xlabel={{\bf \emph{t}:}},
		xmin=0, xmax=17,
		xtick={1,6,11,16},
		xmajorgrids,
		ylabel={\bf Average degree},
		every axis x label/.style={
			at={(-0.06,-0.07)}
		},
		legend image post style={scale=0.8},
		ymajorgrids,
		axis lines*=left,
		xlabel={{\bf \emph{t}:}},
		every axis x label/.style={
			at={(-0.05,-0.095)}
		},
		ymajorticks=true,
		legend style ={ at={(0.47,1.26)}, 
			legend columns=2,
			font=\scriptsize,
			only marks,
			/tikz/every even column/.append style={column sep=0.2cm},
			anchor=north , draw=none, 
			fill=white,align=left},
		cycle list name=black white,
		smooth
		]
		
		\addplot[,mark=square*,mark options={solid, fill=red}] coordinates{
			(1,3.07)
			(2,3.29)
			(3,3.53)
			(4,3.56)
			(5,4.16)
			(6,4.39)
			(7,4.05)
			(8,4.21)
			(9,4.72)
			(10,5.38)
			(11,5.12)
			(12,5.31)
			(13,5.74)
			(14,5.87)
			(15,5.81)
			(16,6.06)
		};
		\addlegendentry{Authenticus};
		
		\addplot[,mark=*,mark options={solid, fill=red}] coordinates{
			(10,6.90)
			(11,15.09)
			(12,35.99)
			(13,49.23)
			(14,50.97)
			(15,49.87)
			(16,53.20)
		};
		\addlegendentry{arxiv};
		\end{axis}
		\end{tikzpicture}%
	\end{subfigure}
	\begin{subfigure}{.30\linewidth}\centering
		\begin{tikzpicture}
		\begin{axis}[
		y label style={at={(axis description cs:0.05,.5)}},
		x label style={at={(axis description cs:0.5,.08)}},
		legend image post style={scale=0.8},
		scale only axis,
		width=0.6\linewidth,
		xtick={1,6,11,16},
		xmin=0, xmax=17,
		xmajorgrids,
		ylabel={\textcolor{white}{.}},
		ymajorgrids,
		axis lines*=left,
		xlabel={{\bf \emph{t}:}},
		every axis x label/.style={
			at={(-0.05,-0.095)}
		},
		ymajorticks=true,
		legend style ={ at={(0.47,1.26)}, 
			legend columns=2,
			font=\scriptsize,
			only marks,
			/tikz/every even column/.append style={column sep=0.2cm},
			anchor=north , draw=none, 
			fill=white,align=left},
		cycle list name=black white,
		smooth
		]
		
		\addplot[mark=square*,mark options={solid, fill=blue}] coordinates{
			(1,4.52)
			(2,5.00)
			(3,5.35)
			(4,4.99)
			(5,4.26)
			(6,4.61)
			(7,5.49)
			(8,4.85)
			(9,3.99)
			(10,4.92)
			(11,5.18)
			(12,4.38)
			(13,3.87)
			(14,4.29)
			(15,4.70)
			(16,4.03)
		};
		\addlegendentry{Minneapolis};
		
		\addplot[mark=*,mark options={solid, fill=blue}] coordinates{
			(1,2.570)
			(2,2.590)
			(3,2.793)
			(4,2.769)
			(5,2.427)
			(6,2.615)
			(7,2.170)
			(8,2.204)
			(9,1.967)
			(10,2.101)
			(11,2.192)
			(12,2.079)
			(13,1.862)
			(14,2.190)
			(15,2.124)
			(16,2.118)
		};
		\addlegendentry{Philadelphia};
		\end{axis}
		\end{tikzpicture}%
	\end{subfigure}
	\begin{subfigure}{.30\linewidth}\centering
		\begin{tikzpicture}
		\begin{axis}[
		x label style={at={(axis description cs:0.5,.08)}},
		scale only axis,
		width=0.6\linewidth,
		xtick={1,6,11,16},
		xmin=0, xmax=17,
		legend image post style={scale=0.8},
		xmajorgrids,
		ylabel={\textcolor{white}{.}},
		ymajorgrids,
		axis lines*=left,
		xlabel={{\bf \emph{t}:}},
		every axis x label/.style={
			at={(-0.05,-0.095)}
		},
		ymajorticks=true,
		legend style ={ at={(0.47,1.26)}, 
			legend columns=2,
			font=\scriptsize,
			only marks,
			/tikz/every even column/.append style={column sep=0.2cm},
			anchor=north , draw=none, 
			fill=white,align=left},
		cycle list name=black white,
		smooth
		]
		
		\addplot[mark=square*,mark options={solid, fill=violet}] coordinates{
			(8,21.682)
			(9,19.167)
			(10,20.814)
			(11,17.288)
			(12,14.166)
			(13,15.910)
			(14,14.246)
			(15,14.925)
			(16,16.302)
		};
		\addlegendentry{Emails};
		
		\addplot[mark=*,mark options={solid, fill=violet}] coordinates{
			(1,2.000)
			(2,2.800)
			(3,2.852)
			(4,3.009)
			(5,3.012)
			(6,3.407)
			(7,3.554)
			(8,3.812)
			(9,4.241)
			(10,4.481)
			(11,4.927)
			(12,4.514)
			(13,4.822)
			(14,4.637)
			(15,3.856)
			(16,2.133)
		};
		\addlegendentry{Enron};
		\end{axis}
		\end{tikzpicture}%
	\end{subfigure}
	
	\hspace{-1.2cm}
	\begin{subfigure}{.30\linewidth}\centering
		\vspace{0.08cm}
		\begin{tikzpicture}
		\begin{axis}[
		scale only axis,
		y label style={at={(axis description cs:0.15,.5)}},
		width=0.6\linewidth,
		xlabel={{\bf \emph{t}:}},
		xmin=0, xmax=17,
		xtick={1,6,11,16},
		xmajorgrids,
		ylabel={\bf Average degree},
		every axis x label/.style={
			at={(-0.06,-0.07)}
		},
		legend image post style={scale=0.8},
		ymajorgrids,
		axis lines*=left,
		xlabel={{\bf \emph{t}:}},
		every axis x label/.style={
			at={(-0.05,-0.095)}
		},
		ymajorticks=true,
		legend style ={ at={(0.47,1.46)}, 
			legend columns=2,
			font=\scriptsize,
			only marks,
			/tikz/every even column/.append style={column sep=0.2cm},
			anchor=north , draw=none, 
			fill=white,align=left},
		cycle list name=black white,
		smooth
		]
		
		\addplot[mark=square*,mark options={solid, fill=ForestGreen}] coordinates{
			(1,20.110)
			(2,17.568)
			(3,30.710)
			(4,19.237)
			(5,15.091)
			(6,20.350)
			(7,14.379)
			(8,19.133)
			(9,14.414)
			(10,16.978)
			(11,16.477)
			(12,14.767)
			(13,15.508)
			(14,19.556)
			(15,20.736)
			(16,29.933)
		};
		\addlegendentry{Gallery};
		
		\addplot[mark=*,mark options={solid, fill=ForestGreen}] coordinates{
			(11,12.659)
			(12,12.063)
			(13,15.753)
			(14,10.186)
			(15,12.884)
			(16,10.598)
		};
		\addlegendentry{Conference};
		
		\addplot[mark=triangle*,mark options={solid, fill=ForestGreen}] coordinates{
			(12,14.372)
			(13,16.600)
			(14,14.264)
			(15,14.658)
			(16,13.880)
		};
		\addlegendentry{School};
		
		\addplot[mark=diamond*,mark options={solid, fill=ForestGreen}] coordinates{
			(7,5.222)
			(8,4.343)
			(9,4.169)
			(10,5.314)
			(11,3.323)
			(12,4.324)
			(13,4.377)
			(14,4.638)
			(15,4.647)
			(16,3.032)
		};
		\addlegendentry{Workplace};
		\end{axis}
		\end{tikzpicture}%
	\end{subfigure}
	\begin{subfigure}{.30\linewidth}\centering
		\vspace{0.37cm}
		\begin{tikzpicture}
		\hspace{0.08cm}
		\begin{axis}[
		y label style={at={(axis description cs:0.05,.5)}},
		x label style={at={(axis description cs:0.5,.08)}},
		legend image post style={scale=0.8},
		scale only axis,
		width=0.6\linewidth,
		xtick={1,6,11,16},
		xmin=0, xmax=17,
		xmajorgrids,
		ylabel={\textcolor{white}{.}},
		ymajorgrids,
		axis lines*=left,
		xlabel={{\bf \emph{t}:}},
		every axis x label/.style={
			at={(-0.05,-0.095)}
		},
		ymajorticks=true,
		legend style ={ at={(0.47,1.26)}, 
			legend columns=2,
			font=\scriptsize,
			only marks,
			/tikz/every even column/.append style={column sep=0.2cm},
			anchor=north , draw=none, 
			fill=white,align=left},
		cycle list name=black white,
		smooth
		]
		
		\addplot[mark=square*,mark options={solid, fill=Dandelion}] coordinates{
			(5,2.500)
			(6,2.774)
			(7,2.557)
			(8,2.680)
			(9,2.708)
			(10,2.887)
			(11,3.010)
			(12,2.982)
			(13,3.134)
			(14,3.116)
			(15,3.027)
			(16,2.666)
		};
		\addlegendentry{Escorts};
		
		\addplot[mark=*,mark options={solid, fill=Dandelion}] coordinates{
			(8,1.516)
			(9,1.588)
			(10,1.969)
			(11,2.054)
			(12,2.429)
			(13,2.831)
			(14,3.572)
			(15,4.176)
			(16,6.047)
		};
		\addlegendentry{Twitter};
		\end{axis}
		\end{tikzpicture}%
	\end{subfigure}
	\begin{subfigure}{.30\linewidth}\centering
		\vspace{0.37cm}
		\begin{tikzpicture}
		\hspace{0.23cm}
		\begin{axis}[
		y label style={at={(axis description cs:0.05,.5)}},
		x label style={at={(axis description cs:0.5,.08)}},
		legend image post style={scale=0.8},
		scale only axis,
		width=0.6\linewidth,
		xtick={1,6,11,16},
		xmin=0, xmax=17,
		xmajorgrids,
		ylabel={\textcolor{white}{.}},
		ymajorgrids,
		axis lines*=left,
		xlabel={{\bf \emph{t}:}},
		every axis x label/.style={
			at={(-0.05,-0.095)}
		},
		ymajorticks=true,
		legend style ={ at={(0.47,1.26)}, 
			legend columns=2,
			font=\scriptsize,
			only marks,
			/tikz/every even column/.append style={column sep=0.2cm},
			anchor=north , draw=none, 
			fill=white,align=left},
		cycle list name=black white,
		smooth
		]
		
		\addplot[,mark=square*,mark options={solid, fill=black}] coordinates{
			(1,2.761)
			(2,2.793)
			(3,2.749)
			(4,3.100)
			(5,3.217)
			(6,3.485)
			(7,3.470)
			(8,3.901)
			(9,3.693)
			(10,3.830)
			(11,3.899)
			(12,4.136)
			(13,4.017)
			(14,3.916)
			(15,4.222)
			(16,4.449)
			
		};
		\addlegendentry{Transfers};
		
		\addplot[,mark=*,mark options={solid, fill=SkyBlue}] coordinates{
			(1,0.861)
			(2,1.097)
			(3,1.111)
			(4,1.475)
			(5,1.745)
			(6,2.072)
			(7,2.498)
			(8,3.321)
			(9,3.596)
			(10,3.913)
			(11,3.939)
			(12,3.666)
			(13,3.341)
			(14,3.475)
			(15,3.433)
			(16,4.732)
			
		};
		\addlegendentry{Facebook};
		\end{axis}
		\end{tikzpicture}%
	\end{subfigure}
	\caption{Average degree of the networks by time -- grouped by type.} \label{fig:avgdegree_evol}
	
	\vspace{0.5cm}
	\hspace{-1.2cm}
	\begin{subfigure}{.30\linewidth}\centering
		\begin{tikzpicture}
		\begin{axis}[
		scale only axis,
		y label style={at={(axis description cs:0.15,.5)}},
		width=0.6\linewidth,
		xlabel={{\bf \emph{t}:}},
		xmin=0, xmax=17,
		xtick={1,6,11,16},
		xmajorgrids,
		ylabel={\bf Char. path length},
		every axis x label/.style={
			at={(-0.06,-0.07)}
		},
		legend image post style={scale=0.8},
		ymajorgrids,
		axis lines*=left,
		xlabel={{\bf \emph{t}:}},
		every axis x label/.style={
			at={(-0.05,-0.095)}
		},
		ymajorticks=true,
		legend style ={ at={(0.47,1.26)}, 
			legend columns=2,
			font=\scriptsize,
			only marks,
			/tikz/every even column/.append style={column sep=0.2cm},
			anchor=north , draw=none, 
			fill=white,align=left},
		cycle list name=black white,
		smooth
		]
		
		\addplot[,mark=square*,mark options={solid, fill=red}] coordinates{
			(1,9.06)
			(2,7.69)
			(3,6.68)
			(4,7.23)
			(5,6.73)
			(6,6.39)
			(7,7.06)
			(8,6.67)
			(9,6.26)
			(10,6.00)
			(11,5.85)
			(12,6.07)
			(13,5.68)
			(14,5.58)
			(15,5.44)
			(16,5.31)
		};
		\addlegendentry{Authenticus};
		
		\addplot[,mark=*,mark options={solid, fill=red}] coordinates{
			(10,4.21)
			(11,3.53)
			(12,3.00)
			(13,2.91)
			(14,2.91)
			(15,2.98)
			(16,2.90)
		};
		\addlegendentry{arxiv};
		\end{axis}
		\end{tikzpicture}%
	\end{subfigure}
	\begin{subfigure}{.30\linewidth}\centering
		\begin{tikzpicture}
		\begin{axis}[
		y label style={at={(axis description cs:0.05,.5)}},
		x label style={at={(axis description cs:0.5,.08)}},
		legend image post style={scale=0.8},
		scale only axis,
		width=0.6\linewidth,
		xtick={1,6,11,16},
		xmin=0, xmax=17,
		xmajorgrids,
		ylabel={\textcolor{white}{.}},
		ymajorgrids,
		axis lines*=left,
		xlabel={{\bf \emph{t}:}},
		every axis x label/.style={
			at={(-0.05,-0.095)}
		},
		ymajorticks=true,
		legend style ={ at={(0.47,1.26)}, 
			legend columns=2,
			font=\scriptsize,
			only marks,
			/tikz/every even column/.append style={column sep=0.2cm},
			anchor=north , draw=none, 
			fill=white,align=left},
		cycle list name=black white,
		smooth
		]
		
		\addplot[mark=square*,mark options={solid, fill=blue}] coordinates{
			(1,3.56)
			(2,3.53)
			(3,3.52)
			(4,3.49)
			(5,3.74)
			(6,3.62)
			(7,3.40)
			(8,3.50)
			(9,3.80)
			(10,3.58)
			(11,3.49)
			(12,3.60)
			(13,3.75)
			(14,3.81)
			(15,3.61)
			(16,3.77)
		};
		\addlegendentry{Minneapolis};
		
		\addplot[mark=*,mark options={solid, fill=blue}] coordinates{
			(1,5.740)
			(2,6.040)
			(3,6.009)
			(4,5.356)
			(5,6.016)
			(6,5.770)
			(7,6.574)
			(8,6.857)
			(9,6.007)
			(10,6.138)
			(11,5.171)
			(12,6.348)
			(13,5.806)
			(14,7.097)
			(15,5.503)
			(16,6.537)
		};
		\addlegendentry{Philadelphia};
		\end{axis}
		\end{tikzpicture}%
	\end{subfigure}
	\begin{subfigure}{.30\linewidth}\centering
		\begin{tikzpicture}
		\begin{axis}[
		x label style={at={(axis description cs:0.5,.08)}},
		scale only axis,
		width=0.6\linewidth,
		xtick={1,6,11,16},
		xmin=0, xmax=17,
		legend image post style={scale=0.8},
		xmajorgrids,
		ylabel={\textcolor{white}{.}},
		ymajorgrids,
		axis lines*=left,
		xlabel={{\bf \emph{t}:}},
		every axis x label/.style={
			at={(-0.05,-0.095)}
		},
		ymajorticks=true,
		legend style ={ at={(0.47,1.26)}, 
			legend columns=2,
			font=\scriptsize,
			only marks,
			/tikz/every even column/.append style={column sep=0.2cm},
			anchor=north , draw=none, 
			fill=white,align=left},
		cycle list name=black white,
		smooth
		]
		
		\addplot[mark=square*,mark options={solid, fill=violet}] coordinates{
			(8,1.996)
			(9,1.877)
			(10,1.964)
			(11,2.065)
			(12,2.341)
			(13,2.231)
			(14,2.271)
			(15,2.327)
			(16,2.202)
		};
		\addlegendentry{Emails};
		
		\addplot[mark=*,mark options={solid, fill=violet}] coordinates{
			(1,2.154)
			(2,3.380)
			(3,3.655)
			(4,3.898)
			(5,5.153)
			(6,4.698)
			(7,4.663)
			(8,4.562)
			(9,4.425)
			(10,4.452)
			(11,4.481)
			(12,4.363)
			(13,4.470)
			(14,4.312)
			(15,3.121)
			(16,2.487)
		};
		\addlegendentry{Enron};
		\end{axis}
		\end{tikzpicture}%
	\end{subfigure}
	
	\hspace{-1.2cm}
	\begin{subfigure}{.30\linewidth}\centering
		\vspace{0.08cm}
		\begin{tikzpicture}
		\begin{axis}[
		scale only axis,
		y label style={at={(axis description cs:0.15,.5)}},
		width=0.6\linewidth,
		xlabel={{\bf \emph{t}:}},
		xmin=0, xmax=17,
		xtick={1,6,11,16},
		xmajorgrids,
		ylabel={\bf Char. path length},
		every axis x label/.style={
			at={(-0.06,-0.07)}
		},
		legend image post style={scale=0.8},
		ymajorgrids,
		axis lines*=left,
		xlabel={{\bf \emph{t}:}},
		every axis x label/.style={
			at={(-0.05,-0.095)}
		},
		ymajorticks=true,
		legend style ={ at={(0.47,1.46)}, 
			legend columns=2,
			font=\scriptsize,
			only marks,
			/tikz/every even column/.append style={column sep=0.2cm},
			anchor=north , draw=none, 
			fill=white,align=left},
		cycle list name=black white,
		smooth
		]
		
		\addplot[mark=square*,mark options={solid, fill=ForestGreen}] coordinates{
			(1,2.919)
			(2,2.962)
			(3,2.463)
			(4,3.370)
			(5,3.040)
			(6,2.645)
			(7,3.174)
			(8,2.775)
			(9,3.127)
			(10,3.127)
			(11,2.737)
			(12,3.028)
			(13,3.223)
			(14,2.797)
			(15,2.926)
			(16,2.947)
		};
		\addlegendentry{Gallery};
		
		\addplot[mark=*,mark options={solid, fill=ForestGreen}] coordinates{
			(11,1.946)
			(12,2.175)
			(13,1.985)
			(14,2.363)
			(15,2.159)
			(16,2.204)
		};
		\addlegendentry{Conference};
		
		\addplot[mark=triangle*,mark options={solid, fill=ForestGreen}] coordinates{
			(12,2.901)
			(13,2.767)
			(14,3.012)
			(15,2.809)
			(16,3.075)
		};
		\addlegendentry{School};
		
		\addplot[mark=diamond*,mark options={solid, fill=ForestGreen}] coordinates{
			(7,2.958)
			(8,3.272)
			(9,3.522)
			(10,3.124)
			(11,3.993)
			(12,3.430)
			(13,3.457)
			(14,3.168)
			(15,3.016)
			(16,4.297)
		};
		\addlegendentry{Workplace};
		\end{axis}
		\end{tikzpicture}%
	\end{subfigure}
	\begin{subfigure}{.30\linewidth}\centering
		\vspace{0.37cm}
		\begin{tikzpicture}
		\hspace{0.08cm}
		\begin{axis}[
		y label style={at={(axis description cs:0.05,.5)}},
		x label style={at={(axis description cs:0.5,.08)}},
		legend image post style={scale=0.8},
		scale only axis,
		width=0.6\linewidth,
		xtick={1,6,11,16},
		xmin=0, xmax=17,
		xmajorgrids,
		ylabel={\textcolor{white}{.}},
		ymajorgrids,
		axis lines*=left,
		xlabel={{\bf \emph{t}:}},
		every axis x label/.style={
			at={(-0.05,-0.095)}
		},
		ymajorticks=true,
		legend style ={ at={(0.47,1.26)}, 
			legend columns=2,
			font=\scriptsize,
			only marks,
			/tikz/every even column/.append style={column sep=0.2cm},
			anchor=north , draw=none, 
			fill=white,align=left},
		cycle list name=black white,
		smooth
		]
		
		\addplot[mark=square*,mark options={solid, fill=Dandelion}] coordinates{
			(5,3.800)
			(6,4.711)
			(7,5.658)
			(8,5.995)
			(9,6.133)
			(10,5.847)
			(11,5.737)
			(12,6.057)
			(13,6.080)
			(14,6.085)
			(15,6.203)
			(16,7.170)
		};
		\addlegendentry{Escorts};
		
		\addplot[mark=*,mark options={solid, fill=Dandelion}] coordinates{
			(8,3.102)
			(9,2.986)
			(10,8.183)
			(11,6.907)
			(12,5.755)
			(13,5.708)
			(14,5.572)
			(15,4.959)
			(16,4.512)
		};
		\addlegendentry{Twitter};
		\end{axis}
		\end{tikzpicture}%
	\end{subfigure}
	\begin{subfigure}{.30\linewidth}\centering
		\vspace{0.37cm}
		\begin{tikzpicture}
		\hspace{0.23cm}
		\begin{axis}[
		y label style={at={(axis description cs:0.05,.5)}},
		x label style={at={(axis description cs:0.5,.08)}},
		legend image post style={scale=0.8},
		scale only axis,
		width=0.6\linewidth,
		xtick={1,6,11,16},
		xmin=0, xmax=17,
		xmajorgrids,
		ylabel={\textcolor{white}{.}},
		ymajorgrids,
		axis lines*=left,
		xlabel={{\bf \emph{t}:}},
		every axis x label/.style={
			at={(-0.05,-0.095)}
		},
		ymajorticks=true,
		legend style ={ at={(0.47,1.30)}, 
			legend columns=2,
			font=\scriptsize,
			only marks,
			/tikz/every even column/.append style={column sep=0.2cm},
			anchor=north , draw=none, 
			fill=white,align=left},
		cycle list name=black white,
		smooth
		]
		
		\addplot[,mark=square*,mark options={solid, fill=black}] coordinates{
			(1,5.098)
			(2,5.160)
			(3,5.383)
			(4,4.954)
			(5,4.810)
			(6,4.677)
			(7,4.805)
			(8,4.529)
			(9,4.548)
			(10,4.547)
			(11,4.505)
			(12,4.395)
			(13,4.343)
			(14,4.439)
			(15,4.378)
			(16,4.146)
		};
		\addlegendentry{Transfers};
		
		\addplot[,mark=*,mark options={solid, fill=SkyBlue}] coordinates{
			(1,1.380)
			(2,1.464)
			(3,1.774)
			(4,14.120)
			(5,12.133)
			(6,9.741)
			(7,8.770)
			(8,7.205)
			(9,6.907)
			(10,6.645)
			(11,6.706)
			(12,6.793)
			(13,7.030)
			(14,6.959)
			(15,7.284)
			(16,6.525)
		};
		\addlegendentry{Facebook};
		\end{axis}
		\end{tikzpicture}%
	\end{subfigure}
	\caption{Characteristic path length of the networks by time -- grouped by type.} \label{fig:charpathlen_evol}
\end{figure}

\subsection{Network motifs} \label{sec:motifs}

Both network motifs and graphlets require an exhaustive subgraph census to be performed beforehand (see Problem~\ref{def:genproblem}). In our experiments we performed subgraph census with $k=4$ and $k=5$. Since no significant differences were found between them the results are presented only for the smaller subgraphs. Subgraph enumeration and necessary motif statistical significance tests were performed using \texttt{GT-Scanner}\footnote{\url{http://www.dcc.fc.up.pt/~daparicio/software.html}} by \cite{aparicio2016extending}. Network motifs were enumerated in the final aggregate state of each network from Table~\ref{tab:nets}. This process is prevalent in network motif analysis of static networks \citep{milo2002network,milo2004superfamilies,choobdar2012comparison,wang2014motifs}. Network motifs require a null model in order to assess motif significance, and we use the one proposed by \cite{milo2002network} which generates a set $\mathcal{R}(N)$ of randomized networks that keep the original in- and out-degrees of each node from $N$. Subgraphs are enumerated in $N$ and $\mathcal{R}(N)$ and if a certain subgraph $M_i$ appears with a much higher frequency in $N$ ($Fr(M_i, N)$) than in $\mathcal{R}(N)$ ($<Fr(M_i, \mathcal{R}(N))>$) it is considered a network motif \citep{milo2004superfamilies}. Motif scores $\Delta_i$ are computed for each subgraph $M_i$ (Equation~\ref{eq:motif_score}) and normalized (Equation~\ref{eq:motif_score_normal}). The set of all $\Delta_i$ of $N$ is refereed to as the motif-fingerprint of $N$.

\begin{equation}\label{eq:motif_score}
\Delta_i = \frac{Fr(M_i, N) - <Fr(M_i, \mathcal{R}(N))>}{Fr(M_i, N) + <Fr(M_i, \mathcal{R}(N))>}
\end{equation}

\begin{equation}\label{eq:motif_score_normal}
\Delta_i = \frac{\Delta_i}{\sqrt{\sum {\Delta_{i}}^2}}
\end{equation}

\begin{figure}	
	\centering	
	\includegraphics[width=0.8\linewidth]{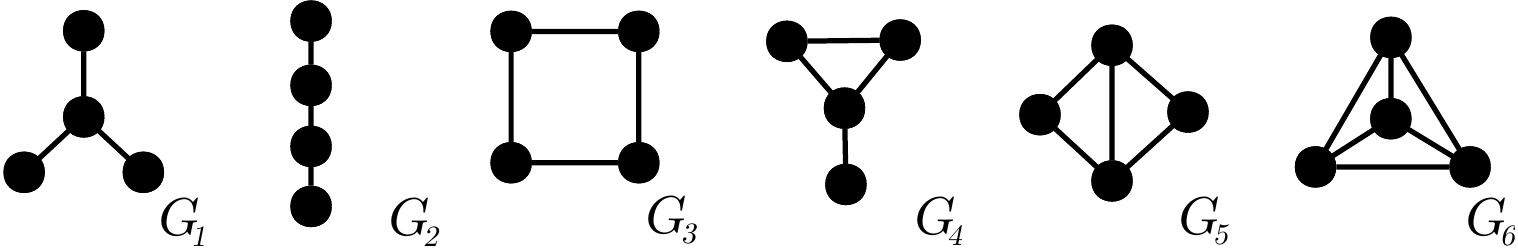}
	
	\hspace{-1.2cm}
	\begin{subfigure}{.30\linewidth}\centering
		\begin{tikzpicture}
		\begin{axis}[
		scale only axis,
		y label style={at={(axis description cs:0.15,.5)}},
		width=0.9\linewidth,
		xtick={1,2,3,4,5,6},
		xticklabels={$\boldsymbol{G_1}$,$\boldsymbol{G_2}$,$\boldsymbol{G_3}$,$\boldsymbol{G_4}$,$\boldsymbol{G_5}$,$\boldsymbol{G_6}$},
		xmajorgrids,
		ymin = -1.1, ymax=1.1,
		ylabel={ $\boldsymbol{\Delta_i}$},
		every axis x label/.style={
			at={(-0.06,-0.07)}
		},
		legend image post style={scale=0.8},
		ymajorgrids,
		axis lines*=left,
		legend style ={ at={(-0.07,1.2)}, 
			legend columns=-1,
			font=\scriptsize,
			anchor=north west, draw=none, 
			fill=white,align=left},
		cycle list name=black white,
		smooth
		]
		
		\addplot[smooth,mark=square*,mark options={solid, fill=red}] coordinates{
			(1,-0.094214704)
			(2,-0.082784356)
			(3,0.023328993)
			(4,0.446500793)
			(5,0.604367508)
			(6,0.647381692)
		};
		\addlegendentry{Authenticus};
		
		\addplot[smooth,mark=*,mark options={solid, fill=red}] coordinates{
			(1,-0.235226990)
			(2,0.101194866)
			(3,-0.323970995)
			(4,0.143828582)
			(5,0.365381401)
			(6,0.569285132)
		};
		\addlegendentry{arxiv};
		\end{axis}
		\end{tikzpicture}%
	\end{subfigure}
	\begin{subfigure}{.30\linewidth}\centering
		\begin{tikzpicture}
		\begin{axis}[
		y label style={at={(axis description cs:0.05,.5)}},
		x label style={at={(axis description cs:0.5,.08)}},
		legend image post style={scale=0.8},
		scale only axis,
		width=0.9\linewidth,
		xtick={1,2,3,4,5,6},
		xticklabels={$\boldsymbol{G_1}$,$\boldsymbol{G_2}$,$\boldsymbol{G_3}$,$\boldsymbol{G_4}$,$\boldsymbol{G_5}$,$\boldsymbol{G_6}$},
		xmajorgrids,
		ymin = -1.1, ymax=1.1,
		ylabel={\textcolor{white}{.}},
		ymajorgrids,
		axis lines*=left,
		legend style ={ at={(-0.18,1.2)}, 
			legend columns=-1,
			font=\scriptsize,
			anchor=north west, draw=none, 
			fill=white,align=left},
		cycle list name=black white,
		smooth
		]
		
		\addplot[mark=square*,mark options={solid, fill=blue}] coordinates{
			(1,0.136370650)
			(2,0.027718435)
			(3,0.628030014)
			(4,-0.286324131)
			(5,-0.208084142)
			(6,-0.678920111)
		};
		\addlegendentry{Minneapolis};
		
		\addplot[mark=*,mark options={solid, fill=blue}] coordinates{
			(1,0.028176427)
			(2,-0.058080020)
			(3,0.661193655)
			(4,-0.545660745)
			(5,-0.220026684)
			(6,-0.460975445)
		};
		\addlegendentry{Philadelphia};
		\end{axis}
		\end{tikzpicture}%
	\end{subfigure}
	\begin{subfigure}{.30\linewidth}\centering
		\begin{tikzpicture}
		\begin{axis}[
		x label style={at={(axis description cs:0.5,.08)}},
		scale only axis,
		width=0.9\linewidth,
		xtick={1,2,3,4,5,6},
		xticklabels={$\boldsymbol{G_1}$,$\boldsymbol{G_2}$,$\boldsymbol{G_3}$,$\boldsymbol{G_4}$,$\boldsymbol{G_5}$,$\boldsymbol{G_6}$},
		xmajorgrids,
		ymin = -1.1, ymax=1.1,
		ylabel={\textcolor{white}{.}},
		ymajorgrids,
		axis lines*=left,
		legend style ={ at={(0.02,1.2)}, 
			legend columns=-1,
			font=\scriptsize,
			anchor=north west, draw=none, 
			fill=white,align=left},
		cycle list name=black white,
		smooth
		]
		
		\addplot[smooth,mark=square*,mark options={solid, fill=violet}] coordinates{
			(1,-0.141562827)
			(2,-0.219322109)
			(3,-0.916544592)
			(4,-0.008764188)
			(5,-0.058661408)
			(6,0.297129295)
		};
		\addlegendentry{Emails};
		
		\addplot[smooth,mark=*,mark options={solid, fill=violet}] coordinates{
			(1,-0.010919133)
			(2,-0.360810129)
			(3,-0.456146074)
			(4,0.082752723)
			(5,0.331971797)
			(6,0.737952773)
		};
		\addlegendentry{Enron};
		\end{axis}
		\end{tikzpicture}%
	\end{subfigure}
	
	\hspace{-1.2cm}
	\begin{subfigure}{.30\linewidth}\centering
		\vspace{0.08cm}
		\begin{tikzpicture}
		\begin{axis}[
		scale only axis,
		y label style={at={(axis description cs:0.15,.5)}},
		width=0.9\linewidth,
		xtick={1,2,3,4,5,6},
		xticklabels={$\boldsymbol{G_1}$,$\boldsymbol{G_2}$,$\boldsymbol{G_3}$,$\boldsymbol{G_4}$,$\boldsymbol{G_5}$,$\boldsymbol{G_6}$},
		xmajorgrids,
		ymin = -1.1, ymax=1.1,
		ylabel={$\boldsymbol{\Delta_i}$	},
		every axis x label/.style={
			at={(-0.06,-0.07)}
		},
		legend image post style={scale=0.8},
		ymajorgrids,
		axis lines*=left,
		legend style ={ at={(-0.10,1.3)}, 
			legend columns=2,
			font=\scriptsize,
			anchor=north west, draw=none, 
			fill=white,align=left},
		cycle list name=black white,
		smooth
		]
		
		\addplot[smooth,mark=square*,mark options={solid, fill=ForestGreen}] coordinates{
			(1,-0.513361575)
			(2,-0.356868016)
			(3,-0.367071691)
			(4,-0.057460979)
			(5,0.271995285)
			(6,0.630143069)
		};
		\addlegendentry{Gallery};
		
		\addplot[smooth,mark=*,mark options={solid, fill=ForestGreen}] coordinates{
			(1,-0.393792363)
			(2,-0.146839706)
			(3,-0.712523579)
			(4,0.003036211)
			(5,-0.019017506)
			(6,0.561520205)
		};
		\addlegendentry{Conference};
		
		\addplot[smooth,mark=triangle*,mark options={solid, fill=ForestGreen}] coordinates{
			(1,-0.324848830)
			(2,-0.270426251)
			(3,-0.350064882)
			(4,0.116733170)
			(5,0.414843431)
			(6,0.716293064)
		};
		\addlegendentry{School};
		
		\addplot[smooth,mark=diamond*,mark options={solid, fill=ForestGreen}] coordinates{
			(1,-0.249808499)
			(2,-0.202146419)
			(3,-0.408066824)
			(4,0.052650269)
			(5,0.231322388)
			(6,0.820933559)
		};
		\addlegendentry{Workplace};
		\end{axis}
		\end{tikzpicture}%
	\end{subfigure}
	\begin{subfigure}{.30\linewidth}\centering
		\vspace{0.37cm}
		\begin{tikzpicture}
		\hspace{0.08cm}
		\begin{axis}[
		y label style={at={(axis description cs:0.05,.5)}},
		x label style={at={(axis description cs:0.5,.08)}},
		legend image post style={scale=0.8},
		scale only axis,
		width=0.9\linewidth,
		xtick={1,2,3,4,5,6},
		xticklabels={$\boldsymbol{G_1}$,$\boldsymbol{G_2}$,$\boldsymbol{G_3}$,$\boldsymbol{G_4}$,$\boldsymbol{G_5}$,$\boldsymbol{G_6}$},
		xmajorgrids,
		ymin = -1.1, ymax=1.1,
		ylabel={\textcolor{white}{.}},
		ymajorgrids,
		axis lines*=left,
		legend style ={ at={(-0.02,1.2)}, 
			legend columns=-1,
			font=\scriptsize,
			anchor=north west, draw=none, 
			fill=white,align=left},
		cycle list name=black white,
		smooth
		]
		
		\addplot[smooth,mark=square*,mark options={solid, fill=Dandelion}] coordinates{
			(1,0.008218826)
			(2,-0.107232163)
			(3,0.568204983)
			(4,-0.385670201)
			(5,-0.232731218)
			(6,-0.680199594)
		};
		\addlegendentry{Escorts};
		
		\addplot[smooth,mark=*,mark options={solid, fill=Dandelion}] coordinates{
			(1,0.004563489)
			(2,-0.135647781)
			(3,0.308975012)
			(4,-0.465467210)
			(5,-0.447458108)
			(6,-0.685007164)
		};
		\addlegendentry{Twitter};
		\end{axis}
		\end{tikzpicture}%
	\end{subfigure}
	\begin{subfigure}{.30\linewidth}\centering
		\vspace{0.37cm}
		\begin{tikzpicture}
		\hspace{0.23cm}
		\begin{axis}[
		y label style={at={(axis description cs:0.05,.5)}},
		x label style={at={(axis description cs:0.5,.08)}},
		legend image post style={scale=0.8},
		scale only axis,
		width=0.9\linewidth,
		xtick={1,2,3,4,5,6},
		xticklabels={$\boldsymbol{G_1}$,$\boldsymbol{G_2}$,$\boldsymbol{G_3}$,$\boldsymbol{G_4}$,$\boldsymbol{G_5}$,$\boldsymbol{G_6}$},
		xmajorgrids,
		ymin = -1.1, ymax=1.1,
		ylabel={\textcolor{white}{.}},
		ymajorgrids,
		axis lines*=left,
		legend style ={ at={(-0.09,1.2)}, 
			legend columns=-1,
			font=\scriptsize,
			anchor=north west, draw=none, 
			fill=white,align=left},
		cycle list name=black white,
		smooth
		]
		
		\addplot[smooth,mark=square*,mark options={solid, fill=black}] coordinates{
			(1,0.129569785)
			(2,-0.255537816)
			(3,0.231852338)
			(4,-0.499294855)
			(5,-0.344162542)
			(6,-0.704566094)
		};
		\addlegendentry{Transfers};
		
		\addplot[smooth,mark=*,mark options={solid, fill=SkyBlue}] coordinates{
			(1,-0.050465527)
			(2,-0.008133811)
			(3,0.444818270)
			(4,0.496971006)
			(5,0.525230831)
			(6,0.526000163)
		};
		\addlegendentry{Facebook};
		\end{axis}
		\end{tikzpicture}%
	\end{subfigure}
	\caption{Motif-fingerprints of networks $\mathcal{N}$ -- grouped by type.} \label{fig:motifs_evol}	
\end{figure}
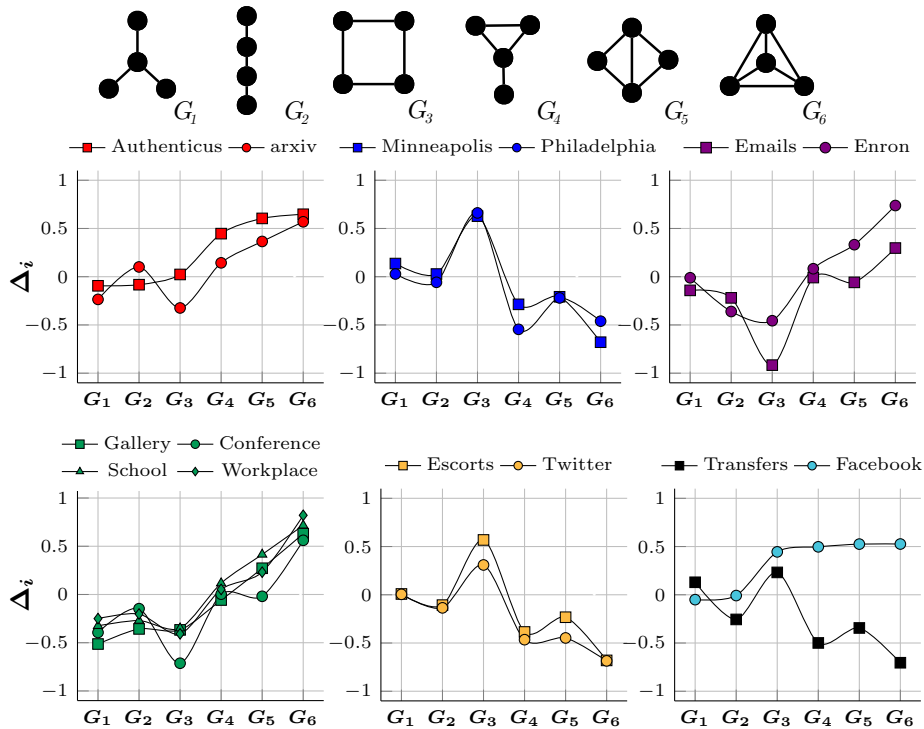

Figure~\ref{fig:motifs_evol} shows the obtained motif-fingerprints for all 4-node undirected subgraphs ($u\mathcal{G}_4$), evaluated against 100 randomized networks. Co-authorship networks have a similar motif-profile where cliques and near-cliques are the most unexpectedly prevalent groups. This comes from the fact that scientific collaboration communities tend to be tightly connected \citep{choobdar2012comparison}. The two crime networks have a similar network profile, with cliques and near-cliques being underrepresented while squares ($G_3$) are very overrepresented. This result was expected since our crime networks are geographical graphs with near-zero clustering coefficient and cities have a grid-like structure. Motif-profiles of the email networks are also relatively alike. Similar to co-authorship networks, cliques and near-cliques are the most overrepresented subgraphs. However, that is much more obvious in \texttt{Enron} than in \texttt{Emails}. This is probably because \texttt{Emails} is too small for the over-representation to become obvious since the small random networks are also capable of generating cliques and near-cliques. Physical interaction networks have a similar motif-fingerprint but it seems indistinguishable from co-authorship networks. Both types of networks have cliques and near-cliques as the most overrepresented subgraphs but those groups have different meanings. In co-authorship networks they might indicate communities but in the short-term event networks they seem to simply indicate that everyone communicates with everyone by the end of the time-frame. Analyzing just the final aggregate network ignores relevant information, it is often more insightful to study how networks evolve. Bipartite networks have similar motif-fingerprints but they are also identical to those of crime networks. It should be pointed out that these networks are not pure bipartite networks but only nearly bipartite, otherwise subgraphs with cycles would never occur ($G_3$, $G_4$, $G_5$ and $G_6$). The \texttt{Transfer} network's motif fingerprint is also similar to the ones of crime and bipartite networks. Finally, \texttt{Facebook}'s motif-profile is alike co-authorship network except $G_3$ is also overrepresented. Since \texttt{Facebook}'s density is so low ($\frac{N}{E^2}\approx\frac{183000}{64000^2} \approx 0.004 \%$) randomized networks have almost exclusively stars ($G_1$) and chains ($G_2$).

\subsection{Static graphlets} \label{sec:graphlets}

Graphlets are subgraphs that take into account the position that nodes occupy in them. Figure~\ref{fig:orbits_u4} shows set $u\mathcal{O}_4$, representing all orbits of $u\mathcal{G}_4$. As stated in Problem~\ref{def:genproblem2}, graphlet-agreement computation requires graphlet-orbits to be counted for all nodes in $N$. After obtaining $GDD$ matrices for all $N \in \mathcal{N}$ we compute the $GDA$ (see Section~\ref{sec:graphlets}) for all network pairs. This results in a $GDA_{i,j}$ matrix where $GDA(N_i, N_j) \approx 0$ means that networks $N_i$ and $N_j$ are completely different and $GDA(N_i, N_j) \approx 1$ translates to $N_i$ and $N_j$ being very similar. Figure~\ref{fig:heatmaps} (a) shows the obtained $GDA_{i,j}$ matrix where each cell is colored according to the $GDA$ value and similar networks have a darker cell. Graphlets group bipartite networks and most of the physical interactions networks correctly. By comparison, using the euclidean distance in order to compare motif-fingerprints two large groups are obtained, as discussed in section~\ref{sec:motifs}. Neither motifs nor graphlets capture temporal information that is necessary to adequately compare temporal networks.  

\begin{figure}
	\includegraphics[width=1.0\linewidth]{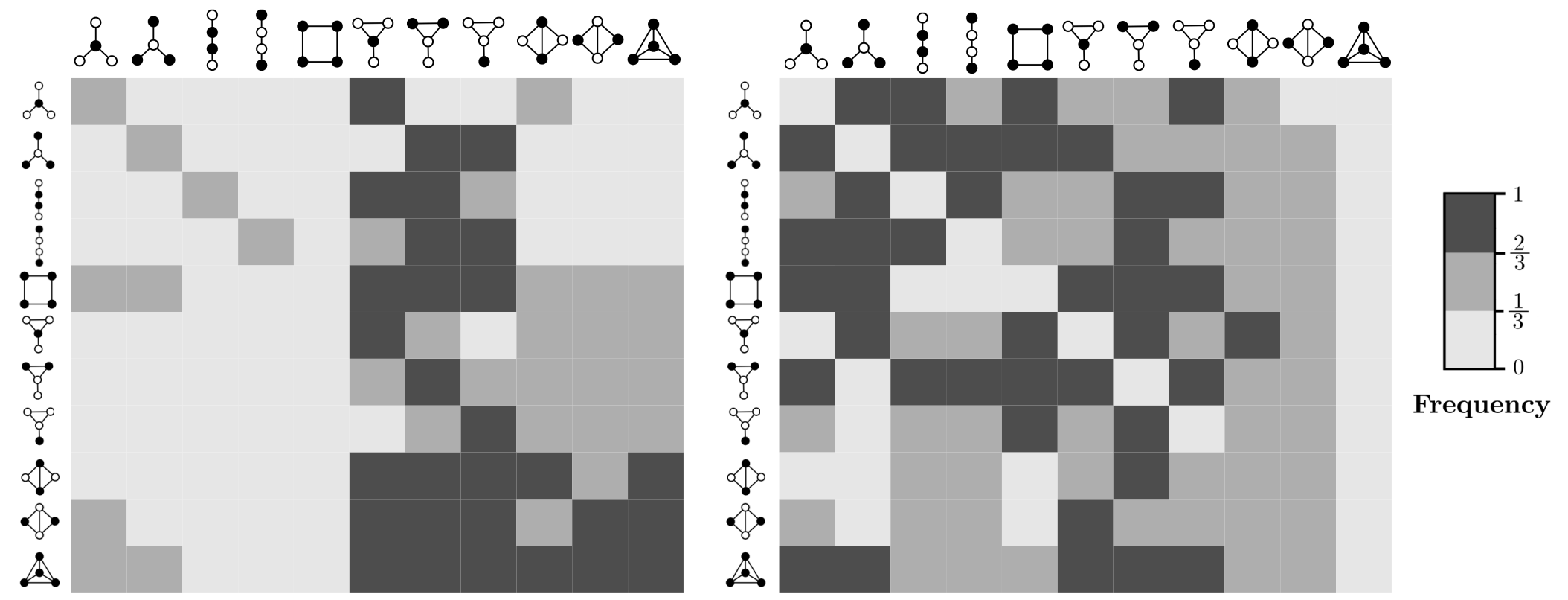}
	\textcolor{white}{.}\hspace{1.5cm} (a) \texttt{Authenticus} \hspace{3.5cm}(b) \texttt{Conference}
	\caption{Orbit transition matrices of (a) a collaboration network and a (b) physical interaction network for all 4-node orbits.}
	\label{fig:authenticus+gallerys_transitions}
\end{figure}

\subsection{Evolving graphlets}

\begin{figure}[h]
	\centering
	\begin{minipage}{0.4\textwidth}
		\includegraphics[width=1.0\linewidth]{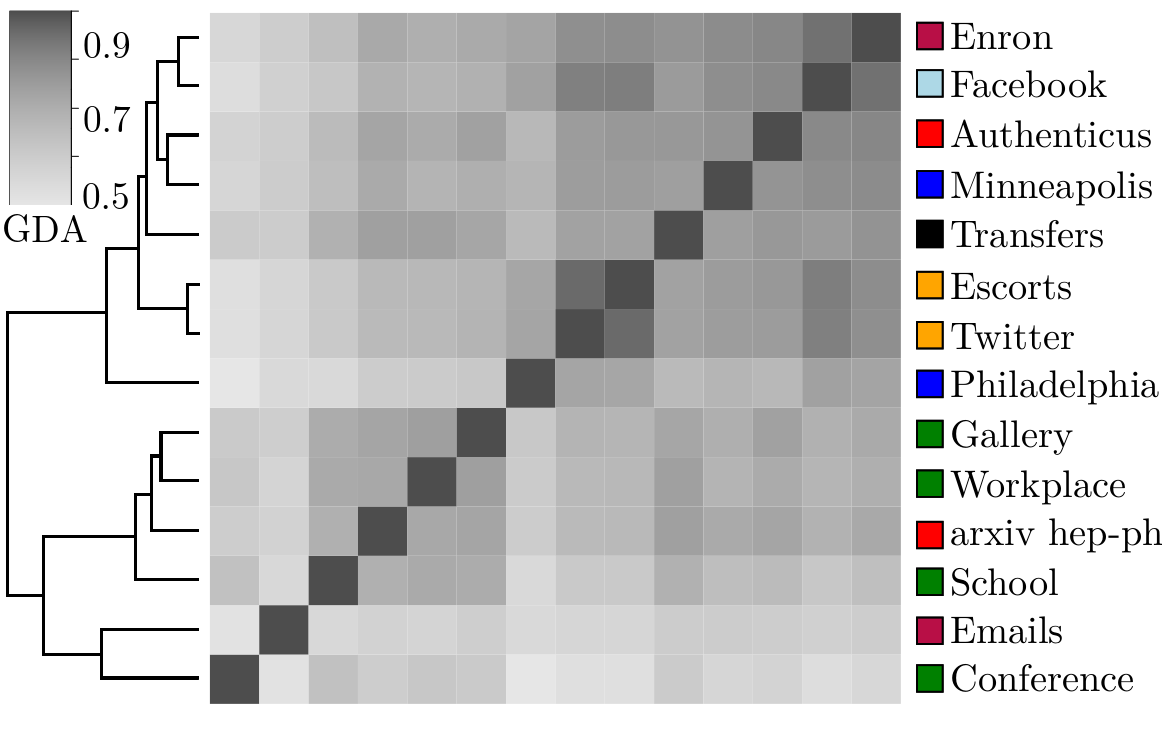}
		\label{fig:gda4_heatmap}
	\end{minipage}
	\begin{minipage}{0.4\textwidth}
		\includegraphics[width=1.0\linewidth]{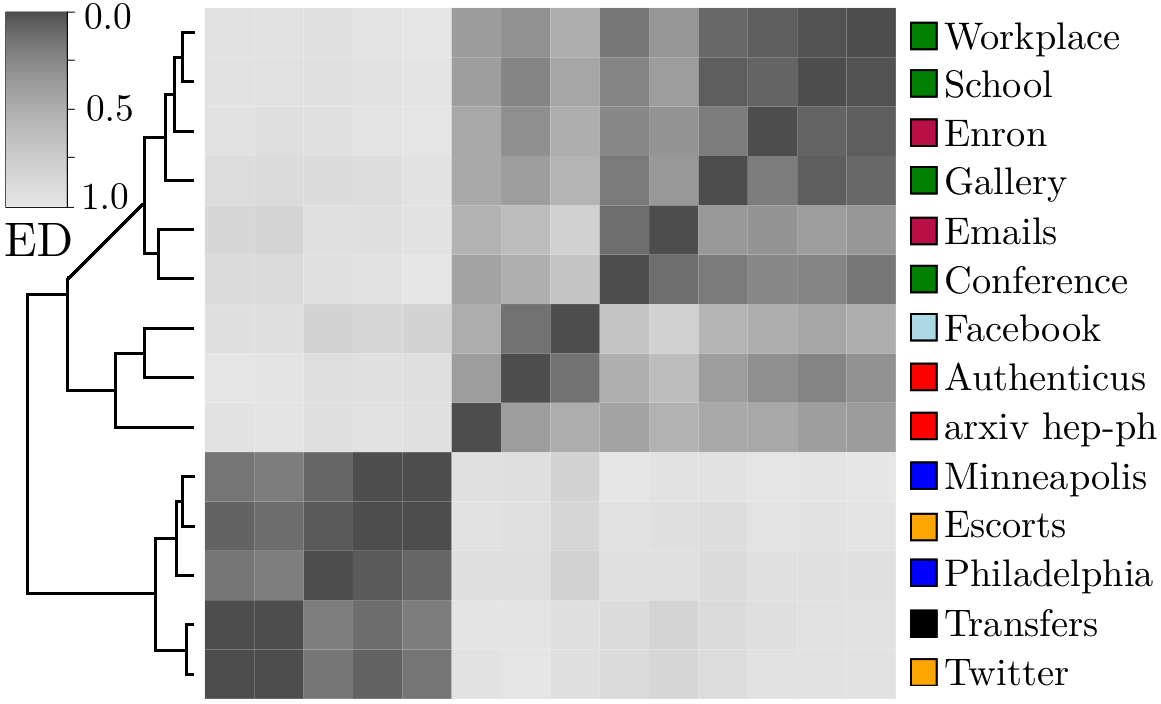}
		\label{fig:motifs4_heatmap}
	\end{minipage}

	\begin{minipage}{0.4\textwidth}
		\vspace{-0.3cm}
		\centering (a)
	\end{minipage}
	\begin{minipage}{0.4\textwidth}
		\vspace{-0.3cm}
		\centering (b)
	\end{minipage}
	
	\begin{minipage}{0.4\textwidth}
		\centering
		\includegraphics[width=1.0\linewidth]{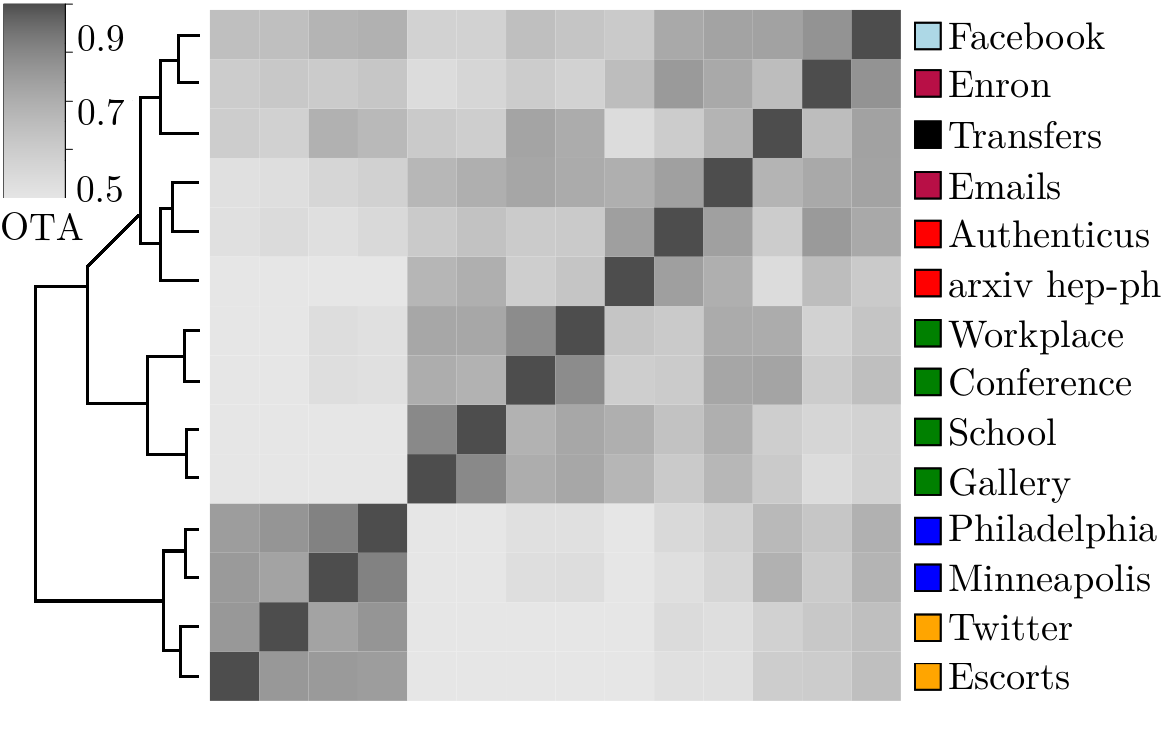}
		\label{fig:ota4_heatmap}
	\end{minipage}
	\begin{minipage}{0.2\textwidth}
		\vspace{-0.1cm}
		\centering (c)
	\end{minipage}
	
	
	\caption{Similarity matrices according to (a) graphlet-degree-agreement ($GDA$), (b) motif-fingerprint distance ($ED$) and (c) orbit-transition-agreement ($OTA$).}
	\label{fig:heatmaps}
\end{figure}

\begin{figure}[h!]
	\centering
	\includegraphics[width=1.0\linewidth]{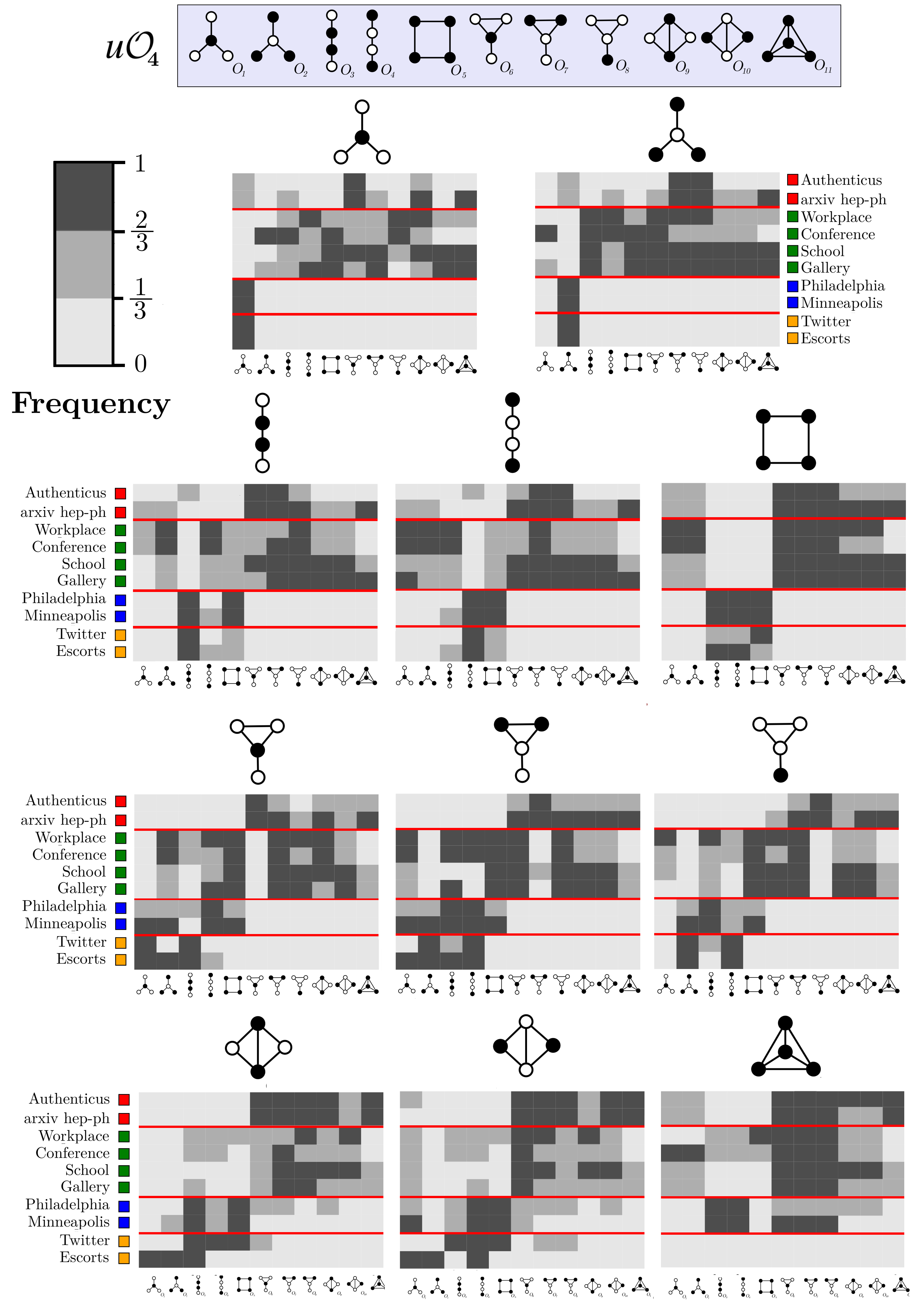}
	\caption{Orbit-transition fingerprints for collaboration, physical interaction, crime and bipartite networks. Frequency values are discretized into \emph{rare}, \emph{common} and \emph{frequent} transitions. }
	\label{fig:orbit_transitions}
\end{figure}

All possible transitions between $u\mathcal{O}_4$ (Figure~\ref{fig:orbits_u4}) were considered in our experiments. Enumerating larger subgraphs was unnecessary since our method achieves an adequate grouping for $k=4$. Furthermore, larger subgraphs would be harder to visualize. Previous studies analyze transitions of graphlets but not those of orbits \citep{doroud2011evolution, kim2012spatiotemporal}, but the latter give more information. A full enumeration was performed for each snapshot $S_i$ and graphlet-orbit transitions matrices $u\mathcal{T}_4$ were created for every network from Table~\ref{tab:nets}. Figure ~\ref{fig:authenticus+gallerys_transitions} shows the transition matrices of \texttt{Authenticus}, a collaboration network, and \texttt{Conference}, a physical interaction network. For an easier visualization, $OTA$ values were discretized into three intervals, indicating rare ($[0, \frac{1}{3}]$), common ($]\frac{1}{3}, \frac{2}{3}]$) and frequent transitions ($]\frac{2}{3}, 1]$). Observing the matrix diagonal suggests that all orbits are relatively stable in \texttt{Authenticus} except for the square-orbit $O_5$. This is expected from collaboration networks since groups forming a square-graph are only loosely connected, therefore they tend to either get tighter (transition to orbits 6-11) or nearly breaking apart (orbits 1-4). On the other hand, orbits in \texttt{Conference} are very unstable, i.e. they almost always change to another orbit. This is explained by the fact that, in short-term physical interaction groups, connections are mostly temporary and not a strong indicator of community. In this example, people meet in a conference and they might meet people that their "group" already met, but they are mostly interested in meeting more people than establishing strong groups. As another example, orbit-1 shows the effect of hubs in collaboration networks: it is more frequent for hub-like groups to gain a new edge between previously unconnected authors (orbit-6) than for them to remain unconnected. It is also common that not only one but two new edges appear (orbit-9). However, stars (orbit-1 and 2) becoming cliques (orbit-11) is rare in \texttt{Authenticus}. Interestingly, Figure~\ref{fig:orbit_transitions} shows that star-to-clique transitions are common in the other collaboration network, \texttt{arXiv hep-ph}. While \texttt{Authenticus} data covers multiple areas, \texttt{arXiv hep-ph} only has publications pertaining to physicists; therefore, the observed differences may hint that physicists form tighter connections sooner than the average. It also seems that transitions are relatively slow in collaboration networks since it is rare for a loosely connected subgraph to become a densely connected subgraph in just a single jump. The same cannot be said about \texttt{Conference}, where behavior is almost chaotic. These are only some of the possible observations about transition matrices that highlight their interpretive power. For completeness, Figure~\ref{fig:orbit_transitions} presents orbit transitions for collaboration, physical interaction, crime and bipartite network. Matrices are discriminated by starting orbit (each matrix) and by network (each matrix-row) for an easier comparison. It is clear that, while networks of the same category have some differences in their orbit-transition profile, they are more alike than networks from different categories. For instance, the transitions of $O_1$ are clearly distinguish co-authorship from physical interaction networks and both from crime and bipartite networks. However, $O_1$ transitions do not differentiate between crime and bipartite networks. To do so one can look at $O_5$, for instance. Orbit-transition fingerprints are a visual way of interpreting how a network evolves and present much information. Figure~\ref{fig:heatmaps} (c) clearly shows that graphlet-orbit transitions are able to correctly group our set of temporal networks while motifs and static graphlet-orbits could not (Figure~\ref{fig:heatmaps} (a) and \ref{fig:heatmaps} (b)).

\section{Conclusions}\label{sec:conclusions}

In this paper we propose an extension of graphlets for temporal networks and means of comparing them. The effectiveness of our proposed method was assessed in a set of temporal networks with predetermined categories. We began by analyzing how global metrics evolved over time, namely the average-degree, clustering-coefficient and the characteristic path-length. While these metrics give insight into the topological structure of the networks, we find that global metrics are not sufficient to differentiate between categories. Network motif and graphlet analyses were also conducted since they capture richer topological information than aforementioned global metrics. However, since they do not take temporal information into account, they are not adequate for temporal network comparison. Our method adequately clustered the set of networks by category. Furthermore, our method produces highly interpretable results, leading to a better understanding of network evolution and differences between transitions of distinct networks. 


\bibliographystyle{spbasic}      
\bibliography{refs}   

\end{document}